\documentclass[rmp,aps,amsfonts,superscriptaddress,preprintnumbers]{revtex4}
\pdfoutput=1

\input epsf
\usepackage{graphics,epsfig}

\def\beq{\begin{equation}}
\def\eeq{\end{equation}}

\def\d{\delta}
\def\dd{{\rm d}}

\begin{document}

\title{On the Quantum Origin of Structure in the Inflationary Universe}

\author{Daniel Baumann}
\email{dbaumann@princeton.edu}
\affiliation{Department of Physics, Princeton University,
Princeton, NJ 08540 U.S.A.}

\begin{abstract}  
\vskip 7pt
In this lecture I give a pedagogical introduction to inflationary cosmology with a special focus on the quantum generation of cosmological perturbations.
The basic outline of the lecture
is as follows:\\

{PART I.} {\bf Basics of Inflation}: Homogeneous Background Solution\\
\noindent
{1.} Review of Big Bang Cosmology\\
{2.} Shortcomings of the Standard Big Bang Picture\\dical 
{3.} Basics of Inflation\\

{PART II.} {\bf Quantum Generation of Cosmological Perturbations}\\
{1.} Quantum Mechanics of the Simple Harmonic Oscillator\\
{2.} Quantum Fluctuations in de Sitter Space\\

{PART III.} {\bf Inflation - Fact or Fiction?}\\
{1.} What is the Physics of Inflation?\\
{2.} Current Cosmological Observations\\
{3.} Gravitational Waves: The Smoking Gun of Inflation\\

{\it The lecture was first presented to astronomy undergraduate students at Princeton in 2005.}

\end{abstract}                                                                 

\maketitle

\newpage
\tableofcontents


\newpage
\section{INTRODUCTION}

\subsection{Back in Time -- Problems with the Big Bang}

Imagine running the cosmic clock backwards. Expansion of the universe becomes contraction. Galaxies rush toward each other implying that the universe was much smaller in the past. As the universe decreases in size its temperature increases. Matter becomes unstable. Atoms are ripped apart into electrons and nuclei. The universe is an energetic sea of charged particles and light. Going even further back in time we lose control over the physics describing the early universe. The known laws of physics break down. General relativity blows up in our faces as the density and temperature of the universe formally become infinite. At this initial singularity concepts like time and space lose their familiar meaning. In a sense (to be described in more detail below) this state marks the beginning of the universe and the Big Bang emerges from it.

The Hot Big Bang theory makes definite predictions that by now have been confirmed in a spectacular series of observations. There is little doubt that the universe really has emerged from such a high energy state and subsequently cooled by expansion. 
However, there are fundamental problems with this simple picture of the Big Bang.  
It requires the universe to start off in a very finely tuned initial state. Fine-tuning means that the initial state has to be very specially chosen. Any small deviation from that state would lead to a universe dramatically different from ours. Physicists usually become very uncomfortable when they realize that their theories only work if there are setup in a very special and unstable way.

In the 1980s cosmologists proposed a radical solution to this fine-tuning problem \cite{Guth, Linde, Albrecht}. 
They envisioned that the universe expanded exponentially quickly for a fraction of a second very early in its history -- growing from a patch as small as $10^{-26}$ m ({\small $0.00000000000000000000000001$ m}),
one hundred billion times smaller than a proton, to macroscopic scales on the order of a meter, all within about $10^{-35}$ s ({\small $0.00000000000000000000000000000000001$ s}) -- before slowing down to the more steadily rate of expansion that has characterized the universe ever since. This sounds unbelievable! It is impossible to imagine these time--, length-- and energy--scales. 
How incredible that theorists can characterize such a dramatic era in the history of the universe using only pen and paper!
  
Admittedly, the detailed physics responsible for this rapid burst of expansion is still uncertain, but inflation convincingly solves the problems of the conventional Big Bang scenario and leads to definite predictions that so far have passed every observational test with flying colors.

What is the origin of structure in the universe? Where did galaxies, stars and ultimately life come from? Inflation combined with quantum mechanics provides an elegant mechanism for generating the initial seeds of cosmic structure. It is this beautiful connection between inflationary expansion of the early universe, the probabilistic nature of quantum mechanics and the formation of galaxies that ultimately has convinced most (but surely not all) cosmologists that something like inflation is likely to have occurred fractions of a second after {\it time zero}. 
Physicists have come to appreciate that understanding the behavior of the universe on the largest scales depends crucially on insights about the physics of the very small.

This lecture is devoted to an account of this theory for the universe's dramatic expansion history and the quantum origin of structure.

\subsection{Disclaimer}

These notes were prepared for astronomy undergraduates to complement a lecture I gave in the summer of 2006 as part of the undergraduate research program of the Department of Astrophysical Sciences at Princeton. Everything presented here was covered in a 90 minute lecture, so that many details had to be suppressed and technical aspects were simplified.
This write-up reflects this approach by emphasizing intuitive aspects and physical principles over rigorous mathematics.
By no means is this text intended to be a precise and/or complete review of inflationary cosmology.
For this I refer the reader to the standard reference by Liddle and Lyth \cite{LiddleLyth} or the excellent recent book by Muhkanov \cite{Mukhanov}.  I have drawn further inspiration from Dodelson's book \cite{Dodelson} and the nice article by Hollands and Wald \cite{HollandsWald}.  Although I have recently edited these notes, I have retained the conversational tone of the original write-up.\\

I am particularly grateful to Prof. Malcolm Longair of Cambridge University for finding these notes on my website and encouraging me to make them public.
Elements of my essay will appear in a new edition of his book on galaxy formation \cite{Longair}.

\section{ASPECTS OF STANDARD COSMOLOGY}
\setcounter{equation}{0}
\renewcommand{\theequation}{2.\arabic{equation}}

\subsection{Homogeneous and Isotropic FRW Universe}
Cosmology describes the global structure and evolution of the universe. Assuming
{\it homogeneity} and {\it isotropy} on large scales one is lead to the Friedmann-Robertson-Walker (FRW) metric for the spacetime of the universe:
\beq
\dd s^2=\dd t^2-a^2(t)\left[{\dd r^2\over{1-\kappa r^2}}+
r^2\left(\dd \theta^2+\sin^2\theta \dd\phi^2\right)\right]\, ,
\label{RW}
\eeq
where $\kappa = -1,0,+1$ for an open, flat or closed universe, respectively. $r$ is a {\it comoving} coordinate; the corresponding physical distance is obtained by multiplying with the scale factor $a(t)$, $R=a(t) r$.
Assuming further that general relativity (GR) and the
Einstein equations\footnote{Throughout I am using units where $\hbar = c \equiv 1$ and $8\pi G \equiv M_{\rm pl}^{-2} \equiv 1$.}, $G_{\mu \nu} = T_{\mu \nu}$, are valid on cosmologically large scales and that the matter content of the universe can be parameterized by the energy-momentum tensor for a perfect fluid 
\beq
\label{equ:PF}
T^\mu_{\,\, \nu} = {\rm diag}(\rho,-P,-P,-P)
\eeq
one arrives at the famous {\it Friedmann equations}:
\begin{eqnarray}
H^2 &=& {1\over 3} \rho- {\kappa\over a^2}\, ,
\label{friedmann}
\\
{\ddot{a} \over a} &=& -{1\over 6}\left(\rho+3 P\right)\, ,
\label{friedmann2}
\end{eqnarray}
where I defined the Hubble parameter
\beq	
H \equiv \frac{\dot{a}}{a}\, .
\eeq
Notice, that in an expanding universe ({\it i.e.} ${\dd a \over \dd t} > 0 \ \forall \ t$) filled with ordinary matter ({\it i.e.} matter satisfying the strong energy condition: $\rho+3P \ge 0$) equation (\ref{friedmann2}) implies ${\dd^2 a \over \dd t^2} < 0 \ \forall \ t$. This indicates the existence of a singularity in the finite past: $a(t \equiv 0) = 0$. Of course, this conclusion relies on the assumption that general relativity and the Friedmann equations are applicable up to arbitrarily high energies. Of course, this assumption is almost certainly not true and it is expected that a quantum theory of gravity will resolve the initial big bang singularity.

An immediate consequence of the two Friedmann equations is the {\it continuity equation}\footnote{This may also be derived from $\nabla_\nu T^{\mu \nu} = 0$.}
\beq
{\dd \rho \over \dd t} +3H\left(\rho+P\right)=0\, .
\label{conservation}
\eeq
More heuristically this also follows from the first law of thermodynamics
\begin{eqnarray}
\dd U &=& - P \dd V \nonumber\\
\dd(\rho a^3) &=& -P \dd(a^3) \quad \quad  \Rightarrow \quad \quad \frac{\dd \ln \rho}{\dd \ln a} = -3 (1+w)\, ,\label{cons}
\end{eqnarray}
where I defined the equation of state
\beq
w \equiv \frac{P}{\rho}\, .
\eeq
The conservation equation  (\ref{cons}) 
can be integrated to give
\beq
\rho\propto a^{-3(1+w)}.
\eeq
Together with the Friedmann equation (\ref{friedmann}) this leads to the time evolution of the scale factor
\beq
a \propto t^{2/3(1+w)} \quad \forall \quad w \ne -1\, .
\eeq
In particular we find the following scalings, $a(t)\propto t^{2/3}$, $a(t)\propto t^{1/2}$ and $a(t)\propto \exp(Ht)$, for the scale factor of a flat ($\kappa=0$) universe dominated by non-relativistic matter ($w=0$), radiation or relativistic matter ($w=\frac{1}{3}$) and a cosmological constant ($w=-1$), respectively.
For each species $i$ we define the {\it present} ratio of  the energy density relative to the {\it critical energy density} $\rho_{\rm crit} \equiv 3 H_0^2$
\beq
\Omega_{i} \equiv {\rho_0^{i} \over \rho_{\rm crit}}\, ,
\eeq
and the corresponding equations of state
\beq
w_{i} \equiv {P_{i} \over \rho_{i}}\, .
\eeq
Here and in the following the subscript '0' denotes evaluation of a quantity at the present time $t_0$.
We normalize the scale factor such that $a_0  = a(t_0) \equiv 1$.
This allows one to rewrite the first 
 Friedmann equation  (\ref{friedmann}) as 
\beq
\left({H\over H_0}\right)^2=\sum_i\Omega_{i}
a^{-3(1+w_{i})}+\Omega_\kappa
a^{-2},
\eeq
with $\Omega_\kappa \equiv -\kappa/a_0^2H_0^2$, which implies the following consistency relation
\beq
\sum_i \Omega_{i}+\Omega_\kappa=1.
\eeq
The second Friedmann equation (\ref{friedmann2}) evaluated at $t=t_0$ becomes
\beq
{1 \over a_0 H_0^2} {\dd^2 a_0 \over \dd t^2}=-{1\over 2}\sum_i \Omega_{i}(1+3 w_{i}).
\eeq
Observations of the cosmic microwave background (CMB) and the large-scale structure (LSS) find that the universe is flat ($\Omega_\kappa \sim 0$) and composed of 4\% atoms, 23\% dark matter and 72\% dark energy \cite{Spergel}: $\Omega_b = 0.04$, $\Omega_d = 0.23$, $\Omega_{\Lambda} = 0.72$, with $w_\Lambda \approx -1$.
It is also found that the universe has tiny ripples of
adiabatic, scale-invariant, Gaussian density fluctuations.
What is the physical origin of these primordial perturbations to the homogeneous universe?
In the bulk of this lecture I will describe how quantum fluctuations in the early universe can explain the observed cosmological perturbations.

\subsection{Brief History of the Universe}

The following table [reproduced from Liddle and Lyth \cite{LiddleLyth}]
summarizes key events in the history of the universe and the corresponding time-- and energy--scales:\\

\begin{tabular}{lll}
\hline
\hline
$t$ & $\rho^{1/4}$ & Event\\
\hline\\
$10^{-42}$ s & $10^{18}$ GeV & Inflation begins?\\
$10^{-32\pm 6}$ s & $10^{13 \pm 3}$ GeV &  Inflation ends, Cold Big Bang begins?\\
$10^{-18\pm 6}$ s & $10^{6 \pm 3}$ GeV &   Hot Big Bang begins?\\
$10^{-10}$ s & $100$ GeV &  Electroweak phase transition?\\
$10^{-4}$ s & $100$ MeV &  Quark-hadron phase transition?\\
$10^{-2}$ s & $10$ MeV & $\gamma$, $\nu$, $e$, $\bar{e}$, $n$, and $p$ in thermal equilibrium\\
$1$ s & $1$ MeV & $\nu$ decoupling, $e\bar{e}$ annihilation.\\
$100$ s & $0.1$ MeV  & Nucleosynthesis (BBN)\\
$10^4$ yr & $1$ eV &  Matter-radiation equality\\
$10^5$ yr & $0.1$ eV & Atom formation, photon decoupling (CMB)\\
$\sim 10^9$ yr & $10^{-3}$ eV &  First bound structures form\\
Now & $10^{-4}$ eV (2.73 K) & The present.\\
&& \\
\hline
\hline

\end{tabular}

\section{BASICS OF INFLATION}
\setcounter{equation}{0}
\renewcommand{\theequation}{3.\arabic{equation}}

\subsection{Shortcomings of the Standard Big Bang Picture}
A number of fundamental questions about the universe are raised by the Big Bang theory: 
\begin{enumerate}
\item {Why is the universe geometrical flat on large scales?}
\item {Why is the universe so smooth on large scales?}
\item {What got the Big Bang going?}
\item {What was the origin of the density inhomogeneities \\which eventually grew to from galaxies, stars and planets?}
\end{enumerate}
We first discuss these questions in more detail and then present accelerated expansion of the early universe as a possible resolution of these puzzles.

\subsubsection{Homogeneity Problem}

In the previous section we assumed homogeneity and isotropy of the universe. Why is this a good assumption? This is particularly surprising given that inhomogeneities are gravitationally
unstable and therefore grow with time. Observations of the cosmic microwave background (CMB) 
show that the inhomogeneities were 
much smaller in the past (at last scattering) than today. One thus expects that 
these inhomogeneities were even smaller at earlier times.
How do we explain the smoothness of the early universe?

\subsubsection{Flatness Problem}

Spacetime in general relativity is dynamical, curving in response to matter in the universe. Why then is the universe so closely approximated by flat Euclidean space?
To understand the severity of the problem in more detail consider the
Friedmann equation 
\begin{equation}
H^2 = \frac{1}{3} \rho(a) - \frac{\kappa}{a^2}\, ,
\end{equation}
written as 
\begin{equation}
\label{equ:omega}
\fbox{$\displaystyle
1-\Omega(a) = \frac{-\kappa}{(aH)^2}$}\, ,
\end{equation}
where
\begin{equation}
\Omega(a) \equiv \frac{\rho(a)}{\rho_{\rm crit}(a)}\, , \quad \rho_{\rm crit}(a) \equiv 3 H(a)^2\, .
\end{equation}
Notice that now $\Omega(a)$ is defined to be time-dependent, whereas the $\Omega$'s in the previous sections we constants, $\Omega(a_0)$.
In standard cosmology the comoving Hubble radius, $(aH)^{-1}$,  grows with time and from equation (\ref{equ:omega}) 
 the quantity $|\Omega-1|$ must thus diverge with time. $\Omega=1$ is an {\it unstable fixed point}. Therefore, in standard Big Bang cosmology without inflation, the near-flatness observed today ($\Omega(a_0) \sim 1$)
requires an extreme fine-tuning of $\Omega$ close to $1$ in the early 
 universe.
More specifically, one finds that the deviation from flatness at Big Bang nucleosynthesis (BBN), during the GUT era and at the Planck scale, respectively has to satisfy the following conditions
\begin{eqnarray}
|\Omega(a_{\rm BBN})-1| &\le& {\cal O}(10^{-16})\\
|\Omega(a_{\rm GUT})-1| &\le& {\cal O}(10^{-55})\\
|\Omega(a_{\rm pl})-1| &\le& {\cal O}(10^{-61})\, .
\end{eqnarray}

Another way of understanding the flatness problem is from the following differential equation
\begin{equation}
\label{equ:omega2}
\frac{\dd \Omega}{\dd [\ln a]} = (1+3w) \Omega (\Omega-1)\, .
\end{equation}
Equation (\ref{equ:omega2}) is derived by differentiating (\ref{equ:omega}) and using the continuity equation (\ref{cons}). This makes it apparent that $\Omega=1$ is an unstable fixed point if the strong energy condition is satisfied 
\begin{equation}
\frac{\dd |\Omega-1|}{\dd [\ln a]} > 0 \quad \Longleftrightarrow \quad 1+3w > 0\, .
\end{equation}
Again, why is $\Omega(a_0) \sim {\cal O}(1)$ and not much smaller or much larger?

\subsubsection{Horizon Problem}

I will show in a moment that the
{\it comoving Hubble radius}, $(aH)^{-1}$, characterizes
the fraction of comoving space in causal contact.
During the Big Bang expansion $(a H)^{-1}$ {\it grows} momtonically and the 
{\it fraction of the universe in causal contact increases with time}.
But the near-homogeneity of the CMB tells us that the universe was extremely homogeneous 
at the time of last scattering on a scale encompassing 
 many regions that a priori are causally independent.
How is this possible? 

To be more specific,
consider radial null geodesics in a flat FRW spacetime
\beq
\dd s^2 = \dd t^2 - a(t)^2 \dd r^2 \equiv 0 \quad \quad \Rightarrow \quad \quad \dd r = \pm \frac{\dd t}{a(t)} \equiv \dd \tau\, .
\eeq
We define the
{\it comoving horizon}, $\tau$, as the causal horizon or the maximum distance a light ray can travel between time $0$ and time $t$
\beq
\fbox{$\displaystyle
\tau \equiv \int_0^t \frac{\dd t'}{a(t')} = \int_0^a \frac{\dd a}{Ha^2} = \int_0^a \dd[\ln a] \left(\frac{1}{aH}\right)$}\, .
\eeq
During the standard cosmological expansion the increasing comoving Hubble radius, $\frac{1}{aH}$, is therefore associated with an increasing comoving horizon\footnote{This justifies the common practice of often using the terms 'comoving Hubble radius' and 'comoving horizon' interchangeably. Although these terms should conceptually be clearly distinguished, this inaccurate use of language has become standard.}, $\tau$. In particular for radiation dominated (RD) and matter dominated (MD) universes we find
\begin{eqnarray}
\tau &=& \int_0^a \frac{\dd a}{Ha^2} \nonumber\\
&=& \frac{1}{H_0} \int_0^a \frac{\dd a}{\Omega^{1/2}(a) a^2} \nonumber\\
&\propto& 
\left\{ 
\begin{array}{ll}
a \quad \quad \,\rm{RD}\\
a^{1/2} \quad \rm{MD}
\end{array} \right.  \, . 
\end{eqnarray}
This means that the comoving horizon grows monotonically with time (at least in an expanding universe) which implies that comoving scales entering the horizon today have been far outside the horizon at CMB decoupling.
Why is the CMB uniform on large scales (of order the present horizon)?

\subsubsection{On the Problem of Initial Conditions}

I want to emphasize that the flatness and horizon problems are {\it not} strict inconsistencies in the standard cosmological model. If one assumes that the initial value of $\Omega$ was extremely close to unity and that the universe began homogeneously (but with just the right level of inhomogeneity to explain structure formation) then the universe will continue to evolve homogeneously in agreement with observations. The flatness and horizon problems are therefore really just severe shortcomings in the predictive power of the Big Bang model. The dramatic flatness of the early universe cannot be predicted by the standard model, but must instead be assumed in the initial conditions. Likewise, the striking large-scale homogeneity of the universe is not explained or predicted by the model, but instead must simply be assumed.
 
\subsubsection{What got the Big Bang going?}

Recall the second Friedmann equation
\begin{equation}
{\dd^2 a \over \dd t^2}= -\frac{a}{6} (\rho + 3 P) < 0\, .
\end{equation}
Since ordinary matter ($\rho+3P >0$) can only cause deceleration ($\ddot{a}<0$), what got the Big Bang going?

\subsubsection{A Crucial Idea}

All Big Bang puzzles are solved by a beautifully simple idea:
{\it invert the behavior of the 
comoving Hubble radius} i.e. make it {\it decrease} sufficiently in the very 
early universe. 
The corresponding condition is that 
\beq
\fbox{$\displaystyle
\frac{\dd}{\dd t} \left(\frac{H^{-1}}{a}\right) < 0 \quad \Rightarrow \quad {\dd^2 a \over \dd t^2} >0 \quad \Rightarrow \quad \rho+3P <0 $}\, .
\eeq
In the next section we introduce inflation and discuss its physical origin.

\subsection{Basics of Inflation}

\subsubsection{Conditions for Inflation}  
The three equivalent conditions necessary for inflation are:
\begin{itemize}
\item Decreasing comoving horizon 
\beq
\frac{\dd}{\dd t} \left(\frac{1}{aH}\right) < 0\, .
\eeq
\item Accelerated expansion
\beq
{\dd^2 a \over \dd t^2 } >0\, . 
\eeq
\item Violation of the strong energy condition -- negative pressure
\beq
P < - \frac{1}{3} \rho\, .
\eeq
\end{itemize}

How to do this? Answer: Scalar field with special dynamics! Although no fundamental scalar field has yet been detected in experiments, there are fortunately plenty of such fields in theories beyond the standard model of particle physics.
In fact, in string theory, for example, there are numerous scalar fields (moduli), although it proves very challenging to find just one with the right characteristics to serve as an inflaton candidate. In the following I will therefore describe the dynamics of a generic scalar field leaving the connection with fundamental particle theory for a future revolution in theoretical physics.

\subsubsection{Scalar Field Cosmology}

The dynamics of a scalar field coupled to gravity is governed by the 
action 
\beq
S = \int \dd^4 x \sqrt{-g} \left[\frac{1}{2} R + {\cal L}_\phi \right]\, ,
\eeq
where
\beq
\label{action_scalar_field}
{\cal L}_\phi={1\over 2} g^{\mu \nu} \partial_\mu\phi \, \partial_\nu\phi
-V(\phi).
\eeq
The potential $V(\phi)$ describes the self-interaction of the scalar field.
The energy-momentum tensor for the scalar field matter is given by  
\beq
T_{\mu\nu} \equiv -\frac{2}{\sqrt{-g}} \frac{\delta S_\phi}{\delta g^{\mu \nu}} = \partial_\mu\phi\partial_\nu\phi-g_{\mu\nu}
\left({1\over 2}\partial^\sigma\phi\partial_\sigma\phi
+V(\phi)\right).
\label{Tscalarfield}
\eeq
The field equation of motion is
\beq
\frac{\delta S}{\delta \phi} = \frac{1}{\sqrt{-g}} \partial_\mu (\sqrt{-g} \partial^\mu \phi) + V'(\phi) = 0\, .
\eeq
Assuming the FRW metric (\ref{RW}) for $g_{\mu \nu}$ and restricting oneself to the case of a homogeneous field $\phi(t)$, the scalar energy-momentum tensor takes the form of a perfect fluid (\ref{equ:PF}) with
\begin{eqnarray}
\rho_\phi &=& {1\over 2}\dot\phi^2+V(\phi)\, , \\
P_\phi &=& {1\over 2}\dot\phi^2-V(\phi)\, .
\end{eqnarray}
The resulting equation of state
\beq
w  = \frac{\frac{1}{2} \dot{\phi}^2 -V}{\frac{1}{2} \dot{\phi}^2 +V}
\eeq
shows that a scalar field can lead to negative pressure ($w< 0$) and accelerated expansion ($w<-1/3$) if the potential energy $V$ dominates over the kinetic energy $\frac{1}{2} \dot{\phi}^2$.
The dynamics of the (homogeneous) scalar field 
and the FRW geometry is determined by
\begin{eqnarray}
& &H^2={1\over 3}\left({1\over 2}\dot\phi^2+V(\phi)\right)\, , 
\label{e1}\\
& &\ddot\phi+3H\dot \phi+V_{,\phi}=0\label{e2}\, .
\end{eqnarray}

\subsubsection{Slow-Roll Inflation}
\label{sec:SR}
Inflation occurs if the field is evolving slow enough that the potential energy dominates the kinetic energy, and the second time derivative of $\phi$ is small enough to allow this slow-roll condition to be maintained for a sufficient period. Thus, inflation requires
\begin{eqnarray}
\dot{\phi}^2 &\ll& V(\phi)\, \\
|\ddot{\phi}| &\ll& |3H \dot{\phi}|, |V_{,\phi}|\, .
\end{eqnarray}
Satisfying these conditions requires the smallness of two dimensionless quantities known as {\it slow-roll parameters}
\begin{eqnarray}
\epsilon(\phi) &\equiv& \frac{1}{2} \left(\frac{V_{,\phi}}{V}\right)^2\\
\eta(\phi) &\equiv& \frac{V_{,\phi\phi}}{V}\, .
\end{eqnarray}
In the slow-roll regime 
\beq
\label{equ:SR}
\epsilon, |\eta| \ll 1
\eeq
the background evolution is
\begin{eqnarray}
H^2 &\approx& {1\over 3} V(\phi) \approx \text{const}\, , \\
\dot{\phi} &\approx& -\frac{V_{,\phi}}{3H}\, ,
\end{eqnarray}
and the
spacetime is approximately {\it de Sitter}
\begin{equation}
a(t) \sim e^{Ht}\, , \quad H \approx \text{const}\, .
\end{equation}
To understand the relation between the slow-roll condition (\ref{equ:SR}) and inflation convince yourself that the following is true
\beq
\label{equ:SR2}
\frac{\ddot{a}}{a} = H^2 \left(1+\frac{\dot H}{H^2}\right) \approx H^2 (1-\epsilon)\, .
\eeq
Consequently, if the slow-roll approximation is valid ($\epsilon \ll 1$), then inflation is guaranteed. However, this condition is sufficient but not necessary since the validity of the slow-roll approximation was required to establish the second equality in (\ref{equ:SR2}).

Inflation ends when the slow-roll conditions (\ref{equ:SR}) are violated\footnote{This can be made exact by the use of Hubble slow-roll parameters instead of the potential slow-roll parameters we introduced here (see Appendix B).}
\beq
\epsilon(\phi_{\rm end}) \approx 1\, .
\eeq
The number of $e$-folds before inflation ends is
\begin{eqnarray}
N(\phi) &\equiv& \ln \frac{a_{\rm end}}{a} \nonumber\\
&=&\int_t^{t_{\rm end}} H \dd t \nonumber \\
&\approx& \int_{\phi_{\rm end}}^\phi \frac{V}{V_{,\phi}} \dd \phi \, .
\end{eqnarray}

\subsubsection{Comoving Horizon during Inflation}

The evolution of the comoving horizon is of such crucial importance to the whole idea of inflation that I want to re-emphasize a few important points.\\

Recall the definition of the comoving horizon (= conformal time) as a logarithmic integral of the comoving Hubble radius
\beq
\label{equ:tau}
\fbox{$\displaystyle
\tau = \int_0^a \dd [\ln a'] \ \frac{1}{a' H(a')} $}\, .
\eeq

\begin{figure}[h]
	\centering
		\includegraphics[width=8cm]{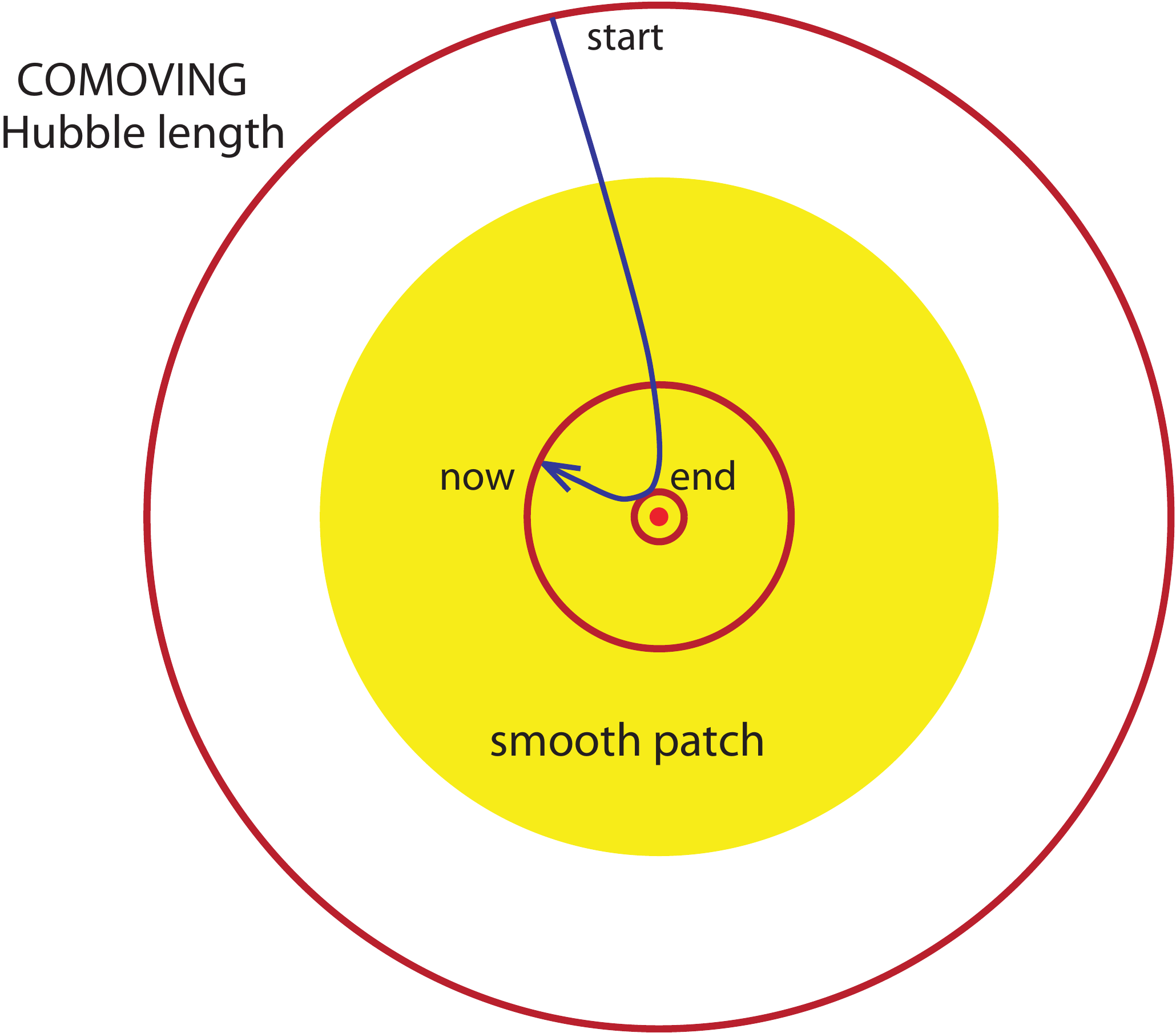}
	\label{fig:horizon}
	\caption{Evolution of the comoving Hubble radius, $(aH)^{-1}$, in the inflationary universe. [Figure adapted from Liddle and Lyth \cite{LiddleLyth}.] The comoving Hubble sphere shrinks during inflation and expands after inflation.}
\end{figure}

Let me emphasize a subtle distinction between the comoving horizon $\tau$ and the comoving Hubble radius $(aH)^{-1}$ (this discussion is from Dodelson \cite{Dodelson}): "If particles are separated by distances greater than $\tau$, they {\it never} could have communicated with one another; if they are separated by distances greater than $(aH)^{-1}$, they cannot talk to each other {\it now}! This distinction is crucial for the solution to the horizon problem which relies on the following: It is possible that $\tau$ is much larger than $(aH)^{-1}$ now, so that particles cannot communicate today but were in causal contact early on. From equation (\ref{equ:tau}) we see that this might happen if the comoving Hubble radius in the early universe was much larger than it is now so that $\tau$ got most of its contribution from early times. Hence, we require a phase of decreasing Hubble radius.
Since $H$ is approximately constant while $a$ grows exponentially during inflation we find that the comoving Hubble radius decreases during inflation just as advertised." \\

Besides solving the Big Bang puzzles the decreasing comoving horizon during inflation is the key feature required for the quantum generation of cosmological perturbations described in the second half of this lecture. I will describe how quantum fluctuations are generated on subhorizon scales, but exit the horizon once the Hubble radius becomes smaller than their comoving wavelength. In physical coordinates this corresponds to the superluminal expansion stretching perturbations to acausal distances. They become classical superhorizon density perturbations which reenter the horizon in the subsequent Big Bang evolution and then gravitationally collapse to form the large scale structure in the universe. 

\subsubsection{Flatness Problem Revisited}

Recall the Friedmann equation (\ref{equ:omega})
\begin{equation}
|1-\Omega(a)| = \frac{1}{(aH)^2}\, .
\end{equation}
Inflation ($H \approx {\rm const.}$, $a = e^{H t}$) is characterized by a decreasing comoving horizon which {\it drives the universe toward flatness} (rather than away from it),
\begin{equation}
|1-\Omega(a)|  \propto \frac{1}{a^2} = e^{-2 H t} \to 0 \quad \text{as} \quad t \to \infty\, .
\end{equation}
This solves the flatness problem! $\Omega=1$ is an attractor during inflation.

 \begin{figure}[h]
	\centering
		\includegraphics[width=9cm]{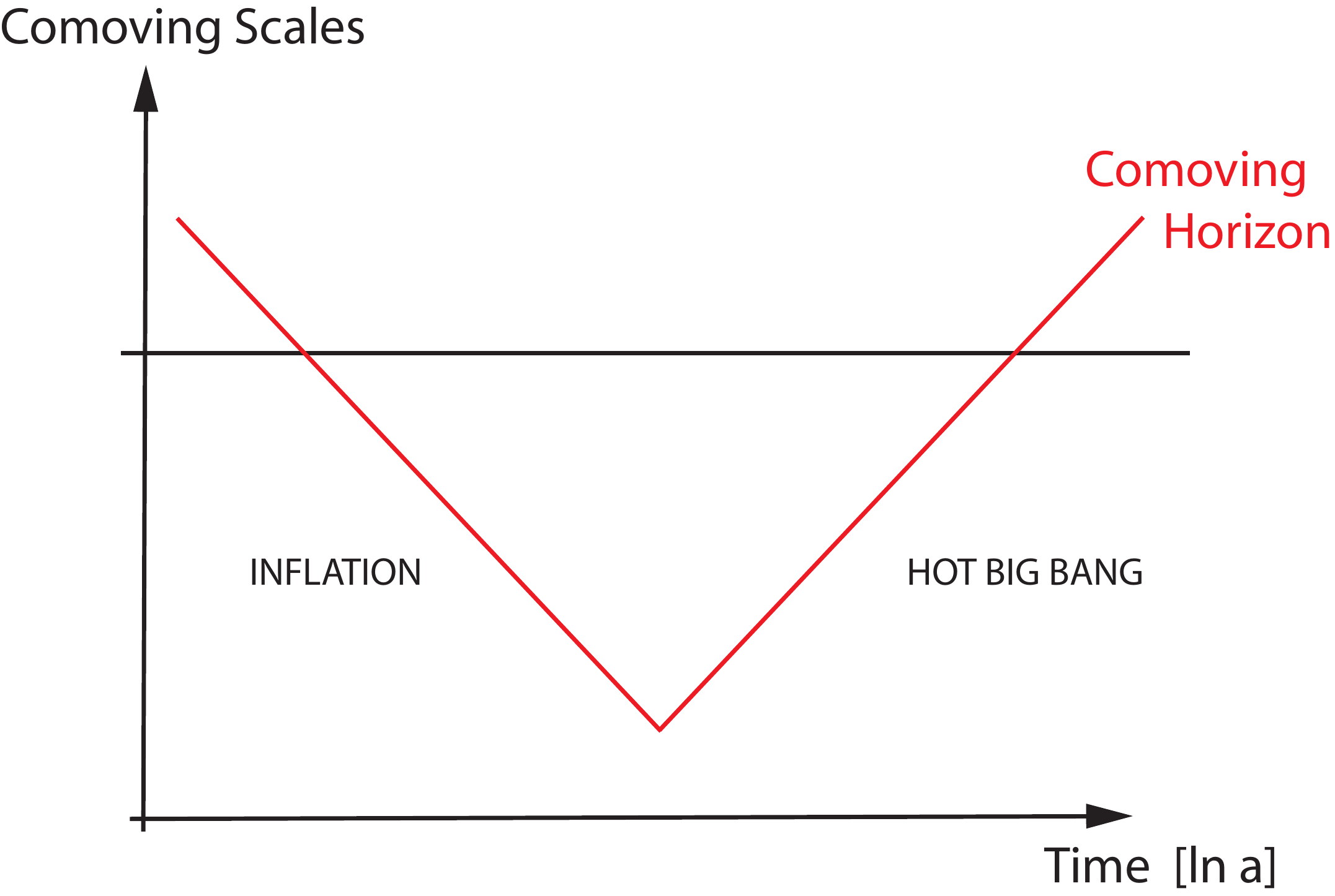}
	\label{fig:horizon}
	\caption{Solution of the Horizon Problem. Scales of cosmological interest were larger than the Hubble radius until $a\sim 10^{-5}$.  However, very early on, before inflation operated, all scales of interest were smaller than the Hubble radius and therefore susceptible to microphysical processing. Similarly, at very late time, scales of cosmological interest came back within the Hubble radius.}
\end{figure}

\subsubsection{Horizon Problem Revisited}

A decreasing comoving horizon means that large scales entering the present universe were inside the horizon before inflation (see Figure 2). Causal physics before inflation therefore established thermal equilibrium and spatial homogeneity. 
The uniformity of the CMB is not a mystery.

\subsubsection{Conformal Diagram of Inflationary Cosmology}

\begin{figure}
	\centering
		\includegraphics[width=9cm]{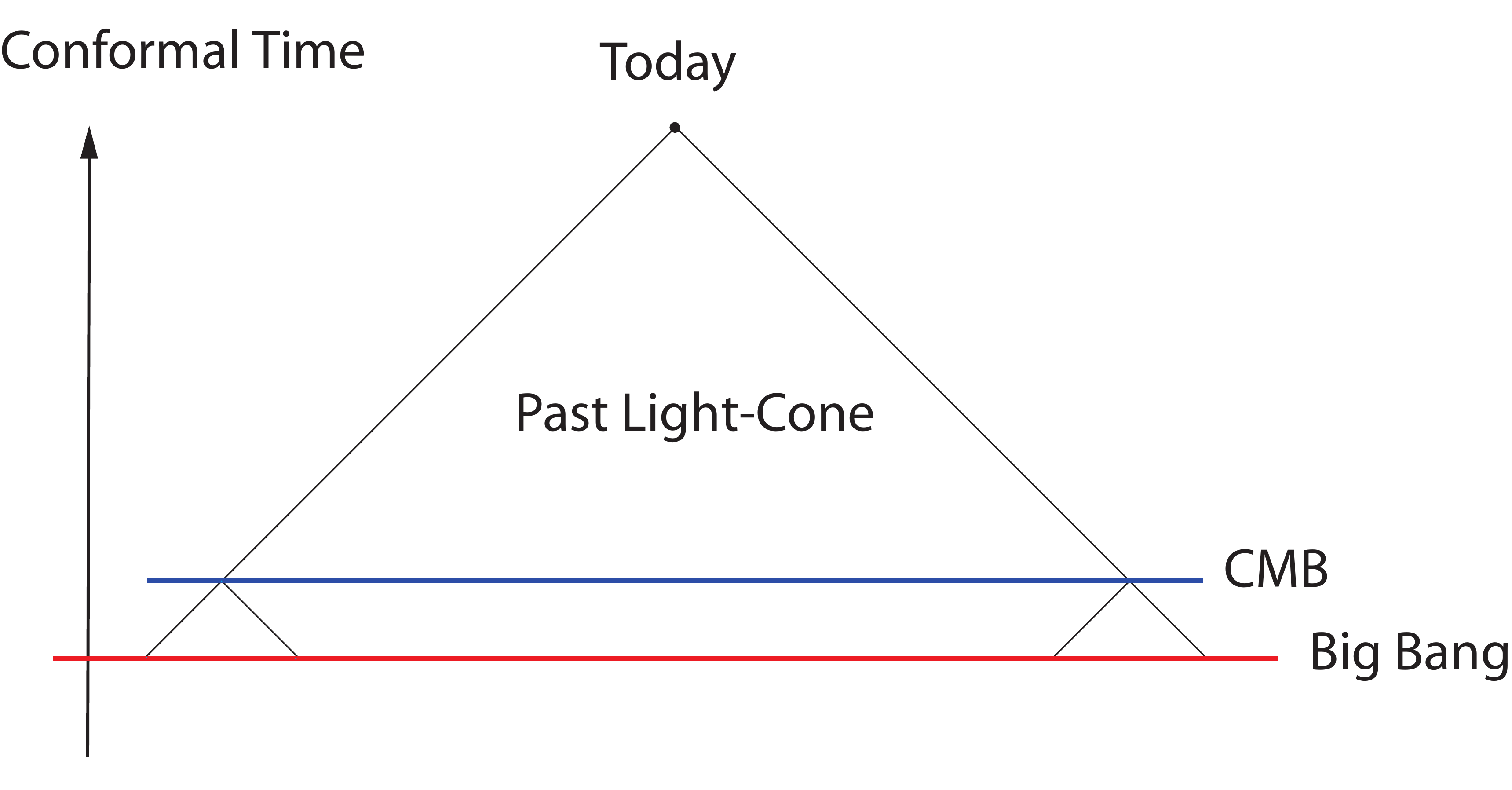}
	\caption{Conformal diagram of Big Bang cosmology. The CMB at last scattering consists of $10^5$ causally disconnected regions!}
	\label{fig:Conformal1b}
\end{figure}
\begin{figure}
	\centering
		\includegraphics[width=9cm]{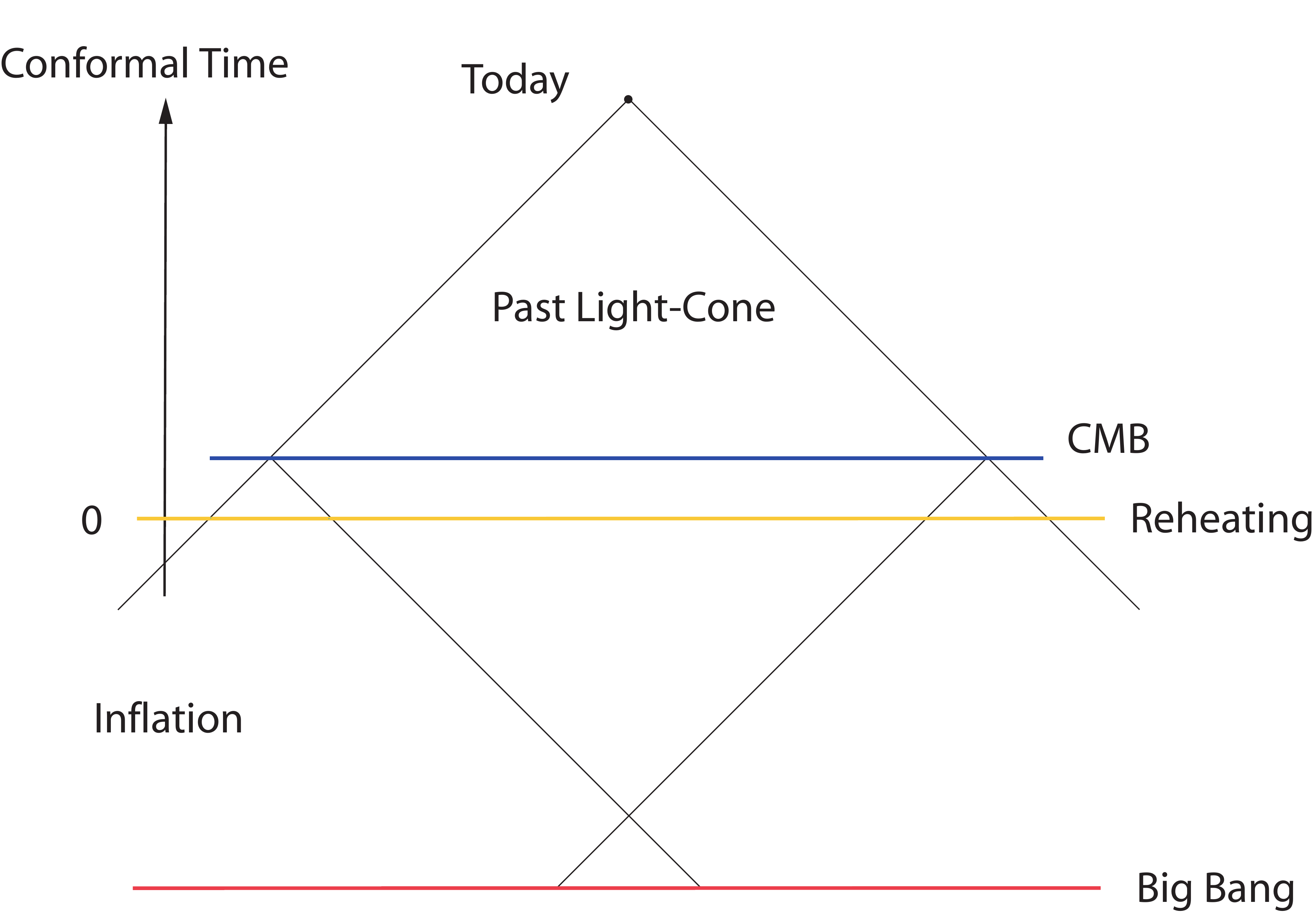}
	\caption{Conformal diagram of inflationary cosmology. Inflation extends conformal time to negative values! The end of inflation creates an "apparent" Big Bang at $\tau = 0$. There is, however, no singularity at $\tau =0$ and the light cones intersect at an earlier time iff inflation lasts for at least 60 $e$-folds.}
		\label{fig:Conformal2}
\end{figure}

A truly illuminating way of visualizing inflation is with the aid of a conformal spacetime diagram.
The flat FRW metric in conformal time $\dd \tau =\dd t/a(t)$ becomes
\beq
\dd s^2 =a^2(\tau) \left[\dd \tau^2 - \dd {\bf x}^2 \right]\, .
\eeq
In conformal coordinates null geodesics ($\dd s^2$\,=\,$0$) are always at $45^\circ$ angles, $\dd \tau = \pm \sqrt{\dd {\bf x}^2} \equiv \pm \dd r$. Since light determines the causal structure of spacetime this provides a nice way to study horizons and thermal equilibrium in inflationary cosmology.

During matter or radiation domination the scale factor evolves as
\beq
a(\tau) \propto
\left\{ 
\begin{array}{ll}
\tau \quad \,\, \text{RD}\\
\tau^2 \quad \text{MD}
\end{array} \right.
\eeq 
Iff the universe had always been dominated by matter or radiation, this would imply the existence of the {\it big bang
singularity} at $\tau = 0$
\beq
a(\tau \equiv 0) = 0\, .
\eeq
The conformal diagram corresponding to standard Big Bang cosmology is given in Figure \ref{fig:Conformal1b}. The horizon problem is apparent. Each spacetime point in the conformal diagram has an associated past light cone which defines its causal past. Two points on a given $\tau = \text{constant}$ surface are in causal contact if their past light cones intersect at the Big Bang, $\tau = 0$. This means that the surface of last scattering ($\tau_{\rm CMB}$) consisted of many causally disconnected regions that won't be in thermal equilibrium. The uniformity of the CMB on large scales hence becomes a serious puzzle.

In de Sitter space, the scale factor is
\beq
\label{equ:dS}
a(\tau) = -\frac{1}{H \tau}\, ,
\eeq
and the singularity, $a=0$, is pushed to the infinite past, $\tau \to -\infty$.
The scale factor  (\ref{equ:dS}) becomes infinite at $\tau = 0$! This is because we have assumed de Sitter space with $H = \text{const.}$, which means that inflation will continue forever with $\tau = 0$ corresponding to the infinite future $t \to +\infty$. In reality, inflation ends at some finite time, and the approximation (\ref{equ:dS}) although valid at early times, breaks down near the end of inflation. So the surface $\tau =0$ is not the Big Bang, but the end of inflation. The initial singularity has been pushed back arbitrarily far in conformal time $\tau \ll 0$, and light cones can extend through the apparent Big Bang so that apparently disconnected points are in causal contact.
This is summarized in the conformal diagram in Figure \ref{fig:Conformal2}.

\subsection{Quantum Fluctuations as the Origin of Structure --
Qualitative Treatment}

A nice heuristic description of inflationary quantum fluctuations can be developed in analogy with a similar treatment of Hawking radiation from black holes.\\

In quantum mechanics the vacuum or empty space is a much more interesting place. Quantum mechanics in fact predicts that empty space is never completely empty. There is always a certain probability that virtual particle--anti-particle pairs pop in and out of existence. Particles and anti-particles are continually created out of nothing, just to annihilate each other a moment later.
In curved space this can result in interesting effects \cite{BD}.

\subsubsection{Hawking Radiation}

Imagine the creation of a particle--anti-particle pair close to the event horizon of a black hole. There is now the possibility that one of the particles crosses the event horizon, the point of no return, and the second particle loses its partner for annihilation and becomes a real particle. This is what an external observer perceives as thermal Hawking radiation coming from the event horizon of the black hole.

\subsubsection{Particle Creation in de Sitter Space } 

An analogous process happens during inflation. 
Spacetime is de Sitter space with constant horizon, $H^{-1}$.
Space expands exponentially, $a(t) \propto e^{Ht}$. Two neighboring points inside the horizion are quickly stretched to a separation much larger than the horizon. Imagine the creation of a particle--anti-particle pair inside the de Sitter horizon. The exponential expansion of space separates them quickly to a distance much greater than the horizon. Being causally disconnected the two particles can't annihilate and become real particles. It is these particles that describe the quantum perturbations of de Sitter space and ultimately become the classical primordial density fluctuations.
The exact mathematically formulation of this process will be the focus of the rest of this lecture.

\section{QUANTUM GENERATION OF COSMOLOGICAL PERTURBATIONS}

\setcounter{equation}{0}
\renewcommand{\theequation}{4.\arabic{equation}}

\subsection{Quantum Mechanics of the Simple Harmonic Oscillator}

Since inflation stretches microscopic scales into astronomical ones, it suggests the possibility that the density perturbations which provide the seeds for galaxy formation might have originated as microscopic quantum fluctuations.
The computation of quantum fluctuations generated by inflation is algebraically quite intensive and it is therefore instructive to start with a simpler example which nevertheless contains most of the relevant physics. Let us therefore warm up by considering the quantization of a one-dimensional simple harmonic oscillator. Harmonic oscillators are one of the few physical systems that physicists know how to solve exactly. Fortunately, almost all more complicated physical systems can be represented by a collection of simple harmonic oscillators with different amplitudes and frequencies (remember normal modes in classical dynamics?). This is of course what Fourier analysis is all about.  
I will show below that free fields in curved spacetime (and de Sitter space in particular) are similar to collections of harmonic oscillators with time-dependent frequencies. The detailed treatment of the quantum harmonic oscillator in this section will therefore not be in vain, but will allow us to shortcut the inflationary calculation.\\

{\sl Action}. The classical action of a harmonic oscillator with time-dependent frequency is
\begin{equation}
S = \int \dd t \, \left(\frac{1}{2} \dot{x}^2 -\frac{1}{2} \omega^2(t) x^2 \right) \equiv \int \dd t \, L\, ,
\end{equation}
where for convenience I have set the particle mass $m \equiv 1$. For concreteness you may wish to consider a particle of mass $m$ on a spring which is heated by an external source so that its spring constant depends on time, $k(t)$, where $\omega^2 =k/m$.\\

{\sl Equation of Motion}. The classical equation of motion follows from variation of the action with respect to the particle coordinate $x$
\begin{equation}
\frac{\delta S}{\delta x} = 0 \quad \Rightarrow \quad \fbox{$\displaystyle \ddot{x} + \omega^2(t)\, x =0 $}\, .
\end{equation}

{\sl Canonical Quantization}. Canonical quantization of the system then proceeds in the standard way:
\begin{itemize}
\item Define the momentum conjugate to $x$
\begin{equation}
p \equiv \frac{\dd L}{\dd \dot{x}} = \dot{x},
\end{equation}
which agrees with the standard notion of particle momentum $p=mv$.
\item Promote the classical variables $x$, $p$ to quantum operators $\hat{x}$, $\hat{p}$ ...
\item ... and impose the canonical commutator
\begin{equation}
\label{equ:commutator}
\fbox{$\displaystyle
\left[\hat{x}, \hat{p} \right] = i \hbar $}\, ,
\end{equation}
where $[\hat x, \hat p] \equiv \hat x \hat p - \hat p \hat x$.
The equation of motion implies that the commutator holds at all times if imposed at some initial time. In particular, for our present example
\begin{equation}
\left[x(t), \dot{x}(t) \right] = i \hbar\, .
\end{equation}
Note that we are in the Heisenberg picture where operators vary in time while states are time-independent.
\item Define the following mode expansion
\begin{equation}
\fbox{$\displaystyle
\hat{x} = v(t) \, \hat{a} + v^*(t) \, \hat{a}^\dagger $}\, ,
\end{equation}
where the (complex) mode function satisfies the classical equation of motion
\begin{equation}
\ddot{v}+\omega^2(t) v =0\, .
\end{equation}
The commutator (\ref{equ:commutator}) becomes
\begin{equation}
\langle v,v \rangle \left[\hat{a}, \hat{a}^\dagger \right]= 1\, ,
\end{equation}
where the bracket notation is defined by
\begin{equation}
\langle v,w \rangle \equiv \frac{i}{\hbar} \left(v^* \partial_t w - (\partial_t v^*) w \right)\, .
\end{equation}
Let us now assume that the solution $v$ is chosen so that the real number $\langle v,v \rangle$ is positive. $v$ can then be rescaled such that $\langle v,v \rangle \equiv 1$ and hence
\begin{equation}
\label{equ:aa}
\fbox{$\displaystyle \left[\hat{a}, \hat{a}^\dagger \right]=1$}\, .
\end{equation}
Equation (\ref{equ:aa}) is
the standard relation for the harmonic oscillator raising and lowering operators.
\item We have hence identified the following annihilation and creation operators
\begin{eqnarray}
\hat{a} &=& \langle v, \hat{x} \rangle\\
\hat{a}^\dagger &=& -\langle v^*, \hat{x} \rangle
\end{eqnarray} 
and can define the vacuum state $| 0 \rangle$ via the prescription
\begin{equation}
\fbox{$\displaystyle \hat{a} | 0 \rangle = 0$}\, .
\end{equation}
Excited states of the system are created by repeated application of creation operators
\begin{equation}
| n \rangle \equiv \frac{1}{\sqrt{n!}} (\hat{a}^\dagger)^n | 0 \rangle\, .
\end{equation}
These states are eigenstates of the number operator $\hat{N} = \hat{a}^\dagger \hat{a}$ with eigenvalue $n$.
\end{itemize}

{\sl Non-uniqueness of the Mode Functions}. So far the solution $v(t)$ is arbitrary, except for the normalization $\langle v,v \rangle =1$. A change in $v(t)$ would be accompanied by a change in $\hat{a} = \langle v, \hat{x}\rangle$ that keeps the solution $x(t)$ unchanged. For the simple harmonic oscillator with time-dependent frequency $\omega(t)$ (and quantum fields in curved spacetime) there is in fact {\it no} unique choice for the mode function $v(t)$. Hence, there is {\it no} unique decomposition of $\hat{x}$ into annihilation and creation operators and {\it no} unique notion of the vacuum. Different choices for the solution $v(t)$ give different vacuum solutions. This problem and its standard (but not uncontested) resolution in the case of inflation will be discussed in more detail below.\\

{\sl Special Case of Constant Frequency $\omega(t) = \omega$}. In the present case we can make progress by considering the special case of a constant frequency harmonic oscillator. In that case a special choice of $v(t)$ is selected if we require that the vacuum state $|0\rangle$ be the ground state of the Hamiltonian. Let's see how this comes about. For general $v$ we have
\begin{eqnarray}
\hat{H} &=& \frac{1}{2} \hat{p}^2 +\frac{1}{2} \omega^2 \hat{x}^2 \nonumber\\
&=& \frac{1}{2} \left[(\dot{v}^2 + \omega^2 v^2) \hat{a} \hat{a}  + (\dot{v}^2 + \omega^2 v^2)^* \, \hat{a}^\dagger \hat{a}^\dagger +(|\dot{v}|^2 +\omega^2 |v|^2) (\hat{a}\hat{a}^\dagger + \hat{a}^\dagger \hat{a})\right]\nonumber\, .
\end{eqnarray} 
Using $\hat{a} |0\rangle =0$ and $[\hat{a},\hat{a}^\dagger]=1$ we hence find the following action of the Hamiltonian operator on the vacuum state
\begin{equation}
\hat{H} |0 \rangle = \frac{1}{2} (\dot{v}^2 + \omega^2 v^2)^*\, \hat{a}^\dagger \hat{a}^\dagger |0 \rangle + \frac{1}{2}(|\dot{v}|^2 +\omega^2 |v|^2) |0 \rangle\, . 
\end{equation}
The requirement that $|0\rangle$ be an eigenstate of $\hat{H}$ means that the first term must vanish which implies the condition
\begin{equation}
\label{equ:modes}
\dot{v} = \pm i \omega v\, .
\end{equation}
For such a function $v$ the norm is
\begin{equation}
\langle v, v \rangle = \mp \frac{2 \omega}{\hbar} |v|^2
\end{equation}
and positivity of the normalization condition $\langle v, v \rangle > 0$ selects the minus sign in equation (\ref{equ:modes})
\begin{equation}
\dot{v} = - i \omega v\, .
\end{equation}
This yields the normalized {\it positive frequency solution} to the equation of motion
\begin{equation}
\label{equ:mode2}
\fbox{$\displaystyle v(t) = \sqrt{\frac{\hbar}{2 \omega}} \, e^{-i \omega t}$}\, .
\end{equation}
With this choice of mode function $v$ the Hamiltonian becomes
\begin{equation}
\hat{H} = \hbar \omega \left(\hat{N}+\frac{1}{2} \right)\, ,
\end{equation}
for which the vacuum $|0\rangle$ is the state of minimum energy $\hbar \omega /2$. If any function other than (\ref{equ:mode2}) is chosen to expand the position operator, then the state annihilated by $\hat{a}$ is {\it not} the ground state of the oscillator.\\

{\sl Zero-point Fluctuations in the Ground State}. 
Consider the mean square expectation value of the position operator $\hat{x}$ in the ground state $|0\rangle$
\begin{eqnarray}
\langle |\hat{x}|^2 \rangle &\equiv& \langle 0|\hat{x}^\dagger \hat{x}|0\rangle \nonumber\\
&=& \langle 0|(v^* \hat{a}^\dagger + v \hat{a})(v \hat{a}+v^* \hat{a}^\dagger )|0\rangle \nonumber\\
&=& |v(\omega,t)|^2 \langle 0|\hat{a} \hat{a}^\dagger |0\rangle \nonumber\\
&=& |v(\omega,t)|^2 \langle 0|[\hat{a}, \hat{a}^\dagger] |0\rangle \nonumber\\
&=& |v(\omega,t)|^2
\end{eqnarray}
This characterizes the  "zero-point fluctuations" of the position in the vacuum state
\begin{equation}
\label{equ:fluc}
\fbox{$\displaystyle \langle |\hat{x}|^2 \rangle = |v(\omega,t)|^2$} = \frac{\hbar}{2\omega}\, .
\end{equation}
This is (almost) all we need to know about quantum mechanics to compute the fluctuation spectrum created by inflation.

\subsection{Quantum Fluctuations in de Sitter Space}

\subsubsection{Fake Quantum Field Theory in Curved Spacetime}

{\sl Disclaimer}. Strictly speaking the following is a "fake calculation", ignoring many important details like perturbations of the background spacetime, backreaction of the inflaton perturbations on the spacetime geometry, subtleties in the vacuum choice, gauge issues, field normalization, etc. Time is short so this is all I can offer for now and I refer to the relevant literature \cite{MFB} or my own notes \cite{proper} for the details. The calculation I am about to present, however, contains all the relevant {\it physics}. If you understand this you will know about the likely physical origin of structure in the universe and may feel inspired to learn the painful details of a rigorous calculation in your own time.

\subsubsection{Harmonic Oscillator Analogy}

Let us first establish the qualitative analogy to the harmonic oscillator. This analogy should be recalled when the going gets rough in the below calculation.
The following argument parallels the discussion in the nice article by Hollands and Wald \cite{HollandsWald}.
\\

We will decompose the inflaton field into a homogeneous background $\phi(t)$ and a perturbation $\d \phi(t, {\bf x})$. $\d \phi$ is analogous to the position coordinate $x$ in our harmonic oscillator example. We will represent $\d \phi$ by its Fourier components $\d \phi_k$ where $k/a \sim \lambda^{-1}$ is the (physical) wave number (inverse wavelength) of each mode. 
I will show below that $\d \phi_k$ satisfies
\beq
\ddot{\d \phi}_k +3 H \dot{\d \phi}_k+\frac{k^2}{a^2} \d \phi_k =0\, .
\eeq
This is identical in form to the harmonic oscillator equation with a unit mass, a (variable) spring constant $k/a$, and a (variable) friction damping coefficient $3H$. 
When the (physical) wavelength, $a/k$, of the mode is much smaller than the Hubble radius, $H^{-1}$, the mode will behave like an ordinary harmonic oscillator, with negligible damping. On the other hand, when the wavelength is much larger than the Hubble radius, the mode will behave like an overdamped oscillator; its "velocity", $\dot{\d \phi}_k$, will rapidly decay towards zero and its amplitude will effectively "freeze".

Both $x$ and $\d \phi$ have calculable zero-point fluctuations in their ground states. For the mechanical harmonic oscillator we just found the scaling $\langle |\hat{x}|^2 \rangle \sim \frac{1}{\omega}$. Similarly, we expect quantum zero-point fluctuations in $\d \phi$ to scale as the inverse (physical) wavenumber since $k/a$ is analogous to the frequency $\omega$ of the harmonic oscillator: $(\d \phi_k)^2 \propto \frac{1}{(k/a)} \propto \lambda$. 

The qualitative picture is therefore the following: 
Quantum fluctuations in $\d \phi$ are generated on subhorizon scales ($k \gg aH$) where they represent oscillating solutions whose amplitude varies as their wavelength. As the mode wavelength is stretched by inflation, the zero-point amplitude increases until the wavelength equals the Hubble radius ($k=aH$) and the amplitude freezes at a constant value as the oscillator becomes overdamped on acausal scales:
\beq
\label{equ:dphik}
(\d \phi_k)^2 \sim \frac{1}{a_\star^3 (k/a_\star)}\, .
\eeq
The normalization factor $a_\star^3$ in (\ref{equ:dphik}) arises from the physical volume element in the Lagrangian for $\phi$ (see Hollands and Wald \cite{HollandsWald}): ${\cal L} \sim \frac{a^3}{2} \dot \phi^2$.
Here, $a_\star$ is the value of the scale factor at the time the mode crossed the Hubble radius
\beq
k=a_\star H_\star\, .
\eeq
Combining the last two expressions we get
\beq
\fbox{$\displaystyle
(\d \phi_k)^2 \sim \frac{H_\star^2}{k^3}$}\, .
\eeq
This is the famous scale-invariant spectrum of inflaton fluctuations which I derive more properly in the following sections.
 It is those frozen inflaton fluctuation modes that are converted into classical density perturbations after reheating and that ultimately lead to the formation of large-scale structure when they reenter the horizon.\\

Let's have fun with the details!

\subsubsection{Action}
Recall the action for a scalar field (minimally) coupled to gravity
\begin{equation}
\label{equ:action}
S = \int \dd^4 x \sqrt{-g} \left[ \frac{1}{2} {\cal R} +\frac{1}{2} (\partial \phi)^2 -V(\phi) \right] \equiv \int \dd^4 x \sqrt{-g} {\cal L}\, ,
\end{equation}
where $\sqrt{-g} = \sqrt{-\det{g_{\mu \nu}}} = a^3(t)$
for the background FRW metric
\begin{equation}
\dd s^2 = \dd t^2 - a^2(t) \dd {\bf x}^2\, .
\end{equation}
The scale factor for de Sitter space is $a(t)=e^{Ht}$. The 
background equation of motion follows from variation of the action
\begin{equation}
\label{equ:KG}
\frac{\delta S}{\delta \phi} =0 \quad \Rightarrow \quad \fbox{$\displaystyle \Box \phi(t, {\bf x}) = \ddot{\phi} - \frac{1}{a^2} \nabla^2 \phi + 3 H \dot{\phi} + V'(\phi)=0$}\, .
\end{equation} 
This is the Klein-Gordon equation of a scalar field in a FRW spacetime.

\subsubsection{Perturbed Equation of Motion}

The inflaton field is then decomposed into a homogeneous background, $\phi(t)$, and a small scale perturbation, $\d \phi(t, {\bf x})$:
\begin{equation}
\phi(t, {\bf x}) = \phi(t) + \delta \phi(t, {\bf x})\, .
\end{equation}
To find the equation of motion for the perturbation $\delta \phi$ we should expand $S$ to {\it second order} in $\delta \phi$ (and include metric perturbations) to get $\delta_2 S$ (have fun!). To avoid this algebraic horror (see Ref.~\cite{MFB}) we instead perturb the homogeneous equation of motion (\ref{equ:KG}) to {\it first order} in $\delta \phi$ (ignoring metric perturbations\footnote{Ignoring metric perturbations is a good approximation for subhorizon scales and can be made exact in {\it spatially flat gauge} 
\beq
\dd s^2=(1+2A)\dd t^2+B_{,i} \dd t \dd x^i + a^2 \dd x^2\nonumber\, .
\eeq
In spatially flat gauge $\d \phi$ decouples from the scalar metric perturbations and our treatment is exact. If you want my treatment implicitly assumes spatially flat gauge or the subhorizon limit.}, remember this is a "fake"!) and then "guess" the second order action $\delta_2 S$.

If the perturbation is expanded in Fourier components
\begin{equation}
\delta \phi(t, {\bf x}) = \int \frac{\dd^3 {\bf k}}{(2\pi)^{3/2}}\,  (\delta \phi_{{\bf k}})\, e^{-i {\bf k} \cdot {\bf x}}
\end{equation}
then the perturbed equation of motion becomes
\begin{equation}
\fbox{$\displaystyle
\ddot{\delta \phi}_{{\bf k}} + 3H \dot{\delta \phi}_{{\bf k}} + \frac{k^2}{a^2} \delta \phi_{{\bf k}} - V_{,\phi \phi} \delta \phi_{{\bf k}} =0 $} \, .
\end{equation}
Imposing the slow-roll condition
\begin{equation}
\eta = \frac{V_{,\phi \phi}}{V} \ll 1 \quad \Rightarrow \quad V_{,\phi \phi} \ll H^2\, ,
\end{equation}
this simplifies to 
\begin{equation}
\ddot{\delta \phi}_{\bf{k}} + 3H \dot{\delta \phi}_{\bf{k}} + \frac{k^2}{a^2} \delta \phi_{\bf{k}} =0 \, .
\end{equation}
The first-order Hubble friction term is then removed by defining the {\it Mukhanov variable}
\begin{equation}
v \equiv a \delta \phi
\end{equation}
and switching to conformal time, $\dd \tau = \dd t/a(t)$,
\begin{equation}
\tau = -\frac{1}{H a}\, .
\end{equation}
We hence arrive at the
master equation of inflationary quantum perturbation theory
\begin{equation}
\label{equ:master}
\fbox{$\displaystyle
v_{\bf{k}}''+\left(k^2- \frac{a''}{a}\right) v_{\bf k}=0 $}\, ,
\end{equation}
where primes denote derivatives with respect to conformal time.
We recognize this as the equation of a
simple harmonic oscillator with time-dependent frequency!

\subsubsection{Action for the Field Perturbation}

The equation of motion (\ref{equ:master}) can be derived from the action
\begin{equation}
\label{equ:d2S}
\delta_2 S = \frac{A}{2} \int \dd \tau \dd^3 {\bf x} \left[(v')^2 -(\partial_i v)^2 + \frac{a''}{a} v^2 \right]\, .
\end{equation}
In this treatment the action is only determined up to an arbitrary multiplicative constant $A$. Had I done the hard work of expanding equation (\ref{equ:action}) to second order in $\delta \phi$, I would have found the same result (\ref{equ:d2S}), but I would also have determined the correct normalization $A=1$.
 
Our starting point for the canonical quantization of fluctuations in the inflaton field will therefore be the action
\begin{equation}
\label{equ:start}
\fbox{$\displaystyle
\delta_2 S = \frac{1}{2} \int \dd \tau \dd^3 {\bf x} \left[(v')^2 -(\partial_i v)^2 + \frac{a''}{a} v^2\right] $}\, .
\end{equation}

\subsubsection{Canonical Quantization of Cosmological Scalar Perturbations}

Canonical quantization of (\ref{equ:start}) proceeds in the standard way: 

\begin{itemize}
\item Define the canonical momentum
\begin{equation}
\pi = \frac{\partial (\delta_2 {\cal L})}{\partial v'} = v'\, .
\end{equation}
\item Promote classical fields to quantum operators
\begin{equation}
v,\pi \quad \Rightarrow \quad \hat{v}, \hat{\pi}\, .
\end{equation}
\item Impose equal time canonical commutation relations
\begin{equation}
[\hat{v}(\tau,{\bf x}), \hat{\pi}(\tau, {\bf x}') ] = i \hbar \, \delta^{(3)}({\bf x}- {\bf x}')\, .
\end{equation}
\item Define the mode decomposition
\begin{equation}
\hat{v}(\tau, {\bf x}) = \int \frac{\dd^3 {\bf k}}{(2\pi)^{3/2}} \left[v_{{\bf k}}(\tau) \hat{a}_{{\bf k}} e^{i {\bf k} \cdot {\bf x}} + \rm{h.c.}\right]\, .
\end{equation}
As usual the mode functions satisfy the classical equations of motion
\begin{equation}
v_{{\bf k}}''+\left(k^2- \frac{a''}{a}\right) v_{\bf k}=0\, ,
\end{equation}
where $a''/a = 2 H^2 a^2 = 2/\tau^2$ for the de Sitter background we are considering.
$\hat{a}_{{\bf k}}$ and $\hat{a}_{{\bf k}}^\dagger$ are annihilation and creation operators with the familiar commutation relation
\begin{equation}
[\hat{a}_{{\bf k}}, \hat{a}_{{\bf k}'}^\dagger] = \delta^{(3)} ({\bf k}-{\bf k}')\, .
\end{equation}
\end{itemize}

\subsubsection{Special Limits of the Solution}

To gain physical intuition it is instructive to study special limits of the equation
\begin{equation}
\label{equ:q}
v_{{\bf k}}''+\left(k^2- \frac{a''}{a}\right) v_{\bf k}=0\, .
\end{equation}

\begin{itemize}
\item {\sl Subhorizon (Minkowski) Limit}.

On subhorizon scales ($k \gg aH$) the $k^2$--term of equation (\ref{equ:q}) dominates over the $\frac{a''}{a}= 2 H^2 a^2$--term. Spacetime is locally static Minkowski space and the equation
\begin{equation}
v_{{\bf k}}''+ k^2 v_{\bf k}=0 \quad \quad \frac{k}{aH} \gg 1\, ,
\end{equation}
has the unique solution of a harmonic oscillator with time-independent frequency (\ref{equ:mode2})
\beq
\label{equ:v1}
v_{{\bf k}} = \frac{e^{-ik \tau}}{\sqrt{2 k}}\, .
\eeq
{Modes oscillate inside the horizon}.

\item {\sl Superhorizon Limit}. 

On superhorizon scales ($k \ll aH$) equation (\ref{equ:q}) becomes 
\begin{equation}
v_{{\bf k}}''- \frac{a''}{a} v_{\bf k}=0\, ,
\end{equation}
whose growing mode solution is
\beq
\label{equ:v2}
v_{{\bf k}} \propto a\, .
\eeq
The scalar field perturbation, $\d \phi$, is therefore found to be constant outside the horizon
\beq
\delta \phi = \frac{v}{a} = \text{const.}
\eeq
{Modes are frozen outside the horizon}.
\end{itemize}

An approximate solution at horizon crossing could now be obtained by matching (\ref{equ:v1}) and (\ref{equ:v2}) at $k=aH$ (WKB approximation). In the present case this is not necessary, since I will now show that de Sitter space in fact permits an exact analytic solution.

\subsubsection{Mode Functions of de Sitter Space}
For the case of a de Sitter background we wish to solve the equation
\begin{equation}
\label{equ:x}
\fbox{$\displaystyle
v_{{\bf k}}''+\left(k^2- \frac{2}{\tau^2}\right) v_{\bf k}=0 $}\, .
\end{equation}
As may be verified by direct substitution
an exact solution to (\ref{equ:x}) is 
\begin{equation}
v_{{\bf k}} = \alpha \, \frac{e^{-i k \tau}}{\sqrt{2k}} \left(1-\frac{i}{k \tau} \right) + \beta \, \frac{e^{i k \tau}}{\sqrt{2k}} \left(1+\frac{i}{k \tau} \right)\, .
\end{equation}
The free parameters $\alpha$ and $\beta$ characterize the non-uniqueness of the mode functions. However, we may fix $\alpha$ and $\beta$ to unique values by considering the subhorizon limit, $k |\tau| \gg 1$, 
\begin{equation}
v_{{\bf k}}''+k^2 v_{\bf k}=0\, .
\end{equation}
This is the equation of a
simple harmonic oscillator with time-independent frequency! 
For this case we know that a unique solution (\ref{equ:mode2}) exists if we require the vacuum to be the minimum energy state. Hence we impose the initial condition
\begin{equation}
\lim_{k\tau \to -\infty} v_{\bf k} = \frac{e^{-ik \tau}}{\sqrt{2 k}}\, . 
\end{equation}
This fixes $\alpha=1$, $\beta=0$ and leads to the unique Bunch-Davies mode functions \cite{BD, Bunch&Davies}
\begin{equation}
\fbox{$\displaystyle
v_{{\bf k}} = \frac{e^{-i k \tau}}{\sqrt{2k}} \left(1-\frac{i}{k \tau} \right) $}\, .
\end{equation}

\subsubsection{Bunch-Davies Vacuum}
The Bunch-Davies vacuum is then defined via
\begin{equation}
\hat{a}_{{\bf k}} |0\rangle =0 \quad \forall \quad {\bf k}\, .
\end{equation}
Note that there is still debate about whether the Bunch-Davies vacuum is the unique inflationary vacuum.
I am not an expert on these issues, so please consult the relevant literature ({\it e.g.} Ref.~\cite{Shenker}) for a more in-depth discussion.

\subsubsection{Super-Horizon Power Spectrum}

Inflationary quantum fluctuations oscillate on scales smaller than the horizon, but freeze on superhorizon scales. Analogous to the case of the simple harmonic oscillator we calculate the zero-point fluctuations of $\d \hat{\phi}$ using the earlier result (\ref{equ:fluc}) that the vacuum expectation value of the square of the field operator is the square of the mode function. In Fourier space the result is
\begin{eqnarray}
\lim_{k \tau \to 0} \langle |\delta \phi_{{\bf k}}|^2 \rangle \equiv \lim_{k \tau \to 0} \langle 0| |\delta \hat{\phi}_{{\bf k}}|^2 |0 \rangle &=& \frac{\lim_{k \tau \to 0} |v_{{\bf k}}|^2}{a^2} \nonumber \\
&=& \frac{H^2}{2 k^3}\, , 
\end{eqnarray}
where I used $\tau = -(a H)^{-1}$ in the last step.
The real space variance of the inflaton perturbation is\footnote{If this was a little fast, here are a few extra steps: Recall the Fourier transform $\d \phi({\bf x}) = \int \dd^3 {\bf k} \d \phi_{{\bf k}} e^{i {\bf k} \cdot {\bf x}}$. We wish to compute
\beq
\langle \d \phi^2 \rangle \equiv \langle \d \phi({\bf x}) \d \phi({\bf x}) \rangle = \int \dd^3 {\bf k} \, \dd^3 {\bf k}'\, \langle \d \phi_{{\bf k}} \d \phi_{{\bf k}'}\rangle\,  e^{i({\bf k}-{\bf k}')\cdot {\bf x}} \, .
\eeq
The delta function in $\langle \d \phi_{{\bf k}} \d \phi_{{\bf k}'}\rangle = \frac{H^2}{2k^3} \delta^{(3)}({\bf k}-{\bf k}')$ allows us to do the ${\bf k}'$-integral to get
\beq
\langle \d \phi^2 \rangle = \int \dd^3 {\bf k} \, \left(\frac{H^2}{2k^3}\right)\, .
\eeq
QED.
}
\begin{eqnarray}
\langle \delta \phi^2 \rangle \equiv \langle 0 | |\d \hat{\phi}|^2 |0\rangle &\sim& \int_{k_i}^{k_f} k^2 \dd k \langle |\delta \hat{\phi}_{{\bf k}}|^2 \rangle \nonumber\, \\
&\sim& \int_{k_i}^{k_f} k^2 \dd k \, \left(\frac{H^2}{k^3}\right) \sim H^2 \ln \left(\frac{k_f}{k_i} \right)\, .
\end{eqnarray}
The integral is cut off at high $k$ because some modes never left the Hubble radius and hence were never amplified, and at low $k$ by the largest wavelengths of interest {\it e.g.} modes which left the Hubble radius just as inflation started.
We have therefore derived the important result
\begin{equation}
\label{equ:dphi}
\fbox{$\displaystyle
\langle \delta \phi^2\rangle \sim H^2$}\, .
\end{equation}

\subsubsection{Time-Delay Formalism}

A very intuitive way of understanding how the quantum fluctuations of the inflaton field $\d \phi$ translate into density fluctuations $\d \rho$ is via the time-delay formalism developed by Guth and Pi \cite{GuthPi}. The basic idea is that $\phi$ controls the time at which inflation ends\footnote{Since we are considering the end of inflation we have of course dropped our assumption of perfect de Sitter space. Assuming an exact de Sitter background served us well in calculating the quantum fluctuations of the inflaton. However, de Sitter space is time-translation invariant so inflation can never end without violating the de Sitter assumption. In fact, the slow-roll parameter
\beq
\epsilon \approx -\frac{\dot{H}}{H^2}
\eeq
parameterizes the deviation from perfect de Sitter space ($H=$const.). During inflation $\epsilon \ll 1$ and the background spacetime is nearly de Sitter. Inflation ends when $\epsilon \approx 1$ and the de Sitter approximation breaks down.}. Regions acquiring a positive frozen fluctuation $\d \phi$ remain potential-dominated longer than regions where $\d \phi$ is negative. Hence, fluctuations of the field $\phi$ lead to a local delay of the time of the end of inflation
\beq
\label{equ:delay}
\fbox{$\displaystyle
\d t = \frac{\d \phi}{\dot{\phi}} $}
\sim \frac{H}{ \dot{\phi}}\, .
\eeq
For the field fluctuation I used the result of the zero-point calculation, $\d \phi = \sqrt{\langle \d \phi^2 \rangle} \sim H $.
After reheating the energy density evolves as $\rho = 3 H^2$ where $H \sim t^{-1}$, so that
\beq
\frac{\d \rho}{\rho} \sim 2 \frac{\d H}{H} \sim H \d t\, .
\eeq
Using (\ref{equ:delay}) and the slow-roll result $3H\dot{\phi} =-V_{,\phi}$ we get
\beq
\label{equ:drho}
\fbox{$\displaystyle
\left(\frac{\d \rho}{\rho}\right)^2 \sim  \frac{V^3}{V'^2} \sim \frac{V}{\epsilon} $}\, .
\eeq
Since $H$ (and hence $V(\phi)$) is not exactly constant during inflation we have to decide when to evaluate the RHS of equation (\ref{equ:drho}). Since perturbations freeze on superhorizon scales it is natural to evaluate the RHS at horizon exit $k=aH$.

\subsubsection{Density Fluctuations}
In describing the density fluctuation spectrum it is conventional to define
\beq
\Bigl\langle \Bigl(\frac{\d \rho}{\rho}\Bigr)^2 \Bigr\rangle \equiv \int \Delta_S^2(k)\,  \dd \ln k\, ,
\eeq
where we have introduced the dimensionless power spectrum
\beq
\Delta_S^2(k) \equiv \frac{k^3 \langle |\delta_k|^2 \rangle}{2 \pi^2}
\eeq
and $\delta_k$ is  the Fourier transform of the fractional density perturbation,
\beq
\delta_{{\bf k}} = \int \frac{\dd^3 {\bf x}}{(2\pi)^{3/2}} \left(\frac{\delta \rho}{\rho}\right) e^{-i {\bf k} \cdot {\bf x}}\, .
\eeq
The dimensionless power spectrum is a function of time, as the amplitude for each mode evolves; it is most common to express the predictions of any specific model in terms of the amplitude of the perturbations at the moment when the physical wavelength of the mode, $\lambda = a/k$, is equal to the Hubble radius, $H^{-1}$,
\beq
\fbox{$\displaystyle
\left.\Delta_S^2(k)\right|_{k=aH} = \left. \frac{1}{24 \pi^2} \frac{V}{\epsilon}\right|_{k=aH}$} \propto k^{n_S -1}\, ,
\eeq
where we have introduced the numerical coefficient resulting from an exact computation.
The scalar spectral index (or tilt) is given by
\beq
\label{equ:tilt}
\fbox{$\displaystyle
n_S-1 \equiv \frac{\dd \ln \Delta_S^2}{\dd \ln k} \approx 2 \eta - 6 \epsilon $}\, .
\eeq
To derive (\ref{equ:tilt}) I used
\beq
\dd \ln k = \frac{\dd k}{k} = \frac{H \dd a}{H a} = \frac{\dot{a}}{a} \dd t = H \dd t = -\frac{H^2}{V'} \dd \phi = -\frac{V}{V'} \dd \phi\, .
\eeq
$n_S=1$ corresponds to a scale-invariant Harrison-Zeldovich-Peebles spectrum.

\subsubsection{Gravitational Waves}

It is not only the nearly-massless inflaton that is excited during inflation, but also any other nearly-massless particle. The other important example is the graviton, which corresponds to tensor perturbations in the metric
\beq
g_{\mu \nu} = g_{\mu \nu}^{(0)} + h_{\mu \nu} \, .
\eeq
$h_{\mu \nu}$ characterizes gravitational waves which are propagating excitations of the gravitational field.
The tensor fluctuation spectrum is derived in direct analogy with the perturbation spectrum of the inflaton perturbation ({\it cf.}~(\ref{equ:dphi})):
\beq
\d h \sim H \sim \sqrt{V}\, ,
\eeq
where $h$ stands for one of the two polarization components of the gravitational wave ($h_+$, $h_\times$).
The tensor fluctuation spectrum is then found to be
\beq
\fbox{$\displaystyle
\left. \Delta_T^2(k)\right|_{k=aH} = \left.\frac{2}{3\pi^2} V \right|_{k=aH} $} \propto k^{n_T}\, .
\eeq
Note that a detection of the tensor spectrum would give us a direct measure of the energy scale of inflation, $V^{1/4}$, and therefore contains important clues on the secrets of the "physics of inflation". 
The tensor spectral index is
\beq
\fbox{$\displaystyle
n_T \equiv \frac{\dd \ln \Delta_T^2}{\dd \ln k} \approx - 2 \epsilon $}\, ,
\eeq
where scale-invariance now corresponds to $n_T =0$ due to an unfortunate historical choice of convention.\\

The tensor-to-scalar ratio is defined as
\beq
\fbox{$\displaystyle
r \equiv \frac{\Delta_T^2}{\Delta_S^2} = 16 \epsilon $}\, .
\eeq

The existence of tensor perturbations is a crucial prediction of inflation that may in principle be verifiable through observations of the polarization of the cosmic microwave background radiation. 

\section{INFLATION -- FACT OR FICTION?}
\setcounter{equation}{0}
\renewcommand{\theequation}{5.\arabic{equation}}



\subsection{What is the Physics of Inflation?}

The inflationary proposal requires a huge extrapolation of the known laws of physics. It is therefore not surprising that the physics governing this phase of superluminal expansion is still very uncertain.
In the absence of a complete theory standard practice has been a phenomenological approach, where an effective potential $V(\phi)$ is postulated. Ultimately, $V(\phi)$ has to be derived from a fundamental theory.

Understanding the (micro)physics of inflation remains one of the most important open problems in modern cosmology and theoretical physics.
Explicit particle physics models of inflation remain elusive.  
A natural microscopic explanation for inflation has yet to be uncovered.
Nevertheless, there have recently been interesting efforts to derive inflation from string theory (see {\it e.g.} \cite{LiamEva} for a review).
Inflation in string theory is still in its infancy, but it seems clear that our understanding of inflation will benefit greatly from a better understanding of moduli stabilization and supersymmetry breaking in string theory.
Hopefully, this will give some insights into which models of inflation are microscopically viable, meaning that they can be derived from explicit string compactifications.
Given the prospect of explicit and controllable models of inflation in string theory, one is led to ask whether these theories have specific observational signatures.  In particular, it will be interesting to explore whether there are predictions that while non-generic in effective field theory, may still have a well-motivated microscopic origin in string theory.

\subsection{Observational Tests of the Inflationary Paradigm}

As of October 2007, the best fit values for the parameters of the inflationary perturbation spectrum are (WMAP+SDSS \cite{Spergel}):
\begin{eqnarray}
\Delta_S(k_0) &=& (4.75 \pm 0.10) \times 10^{-5}\, , \quad k_0 \equiv 0.05 \, \text{Mpc}^{-1}\, ,\\
n_S &=& 0.96 \pm 0.04\, , \quad (r \equiv 0 \ {\rm prior})\, ,\\
r &<& 0.3\, .
\end{eqnarray}
No deviations from the Gaussian and adiabatic predictions have been detected. The universe is flat, $\Omega_\kappa < 0.02$.

\begin{figure}[h]
	\centering
		\includegraphics[width=0.80\textwidth]{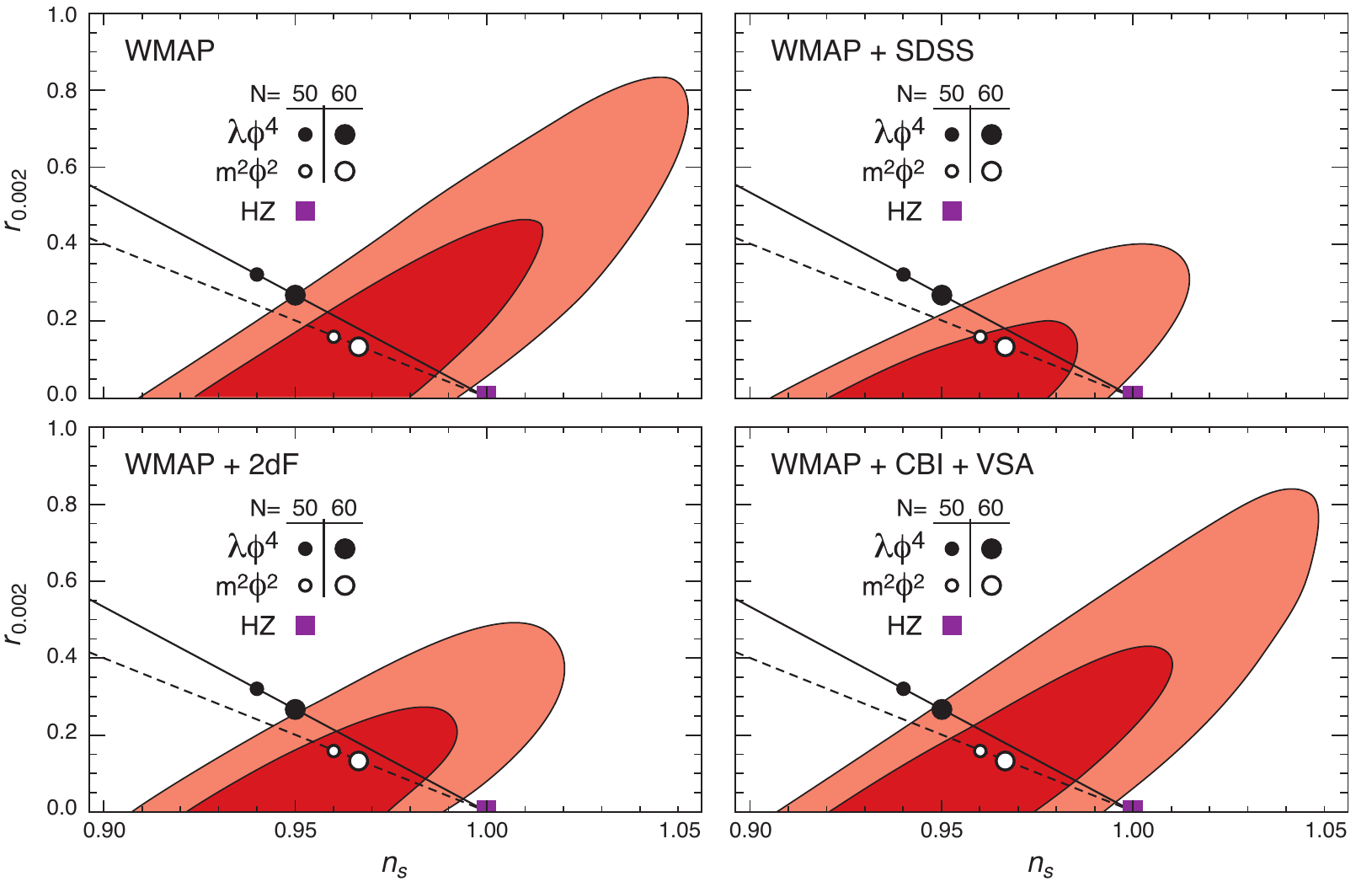}
	\caption{Current constraints on inflationary models in the $n_S-r$ plane. [Reproduced from Ref.~\cite{Spergel}]}
	\label{fig:WMAP}
\end{figure}

The future of cosmological observations is bright.
A number of experiments are planned or already operational that will give detailed information about the basic inflationary parameters. The future of CMB and LSS observations lies in precision measurements of the small scale power spectrum and measurements of CMB polarization.  Measuring $\Delta_S$ on small scales allows a more precise determination of the important parameter $n_S$ and test for its scale-dependence of running $\frac{\dd n_S}{\dd \ln k}$.  In addition, future CMB and LSS observations will place the first meaningful constraint on the non-Gaussianity of the primordial power spectrum, $f_{NL}$. Finally, B-modes of CMB polarization provides the unique way to test for a primordial background of tensor perturbations, $r$.
A future detection of B-modes will be a smoking gun signature of inflation.


\subsection{Gravitational Waves: The Smoking Gun of Inflation}

A conservative estimate for a gravitational wave amplitude that is accessible to future experiments is $r > 0.01$.
Any signal that is much smaller than this will forever be drowned in tensor signals from astrophysical foregrounds.\\

Two pieces of information follow from  a measurement of the primordial tensor-to-scalar ratio $r$:
\begin{enumerate}
\item Energy Scale of Inflation\\
The measured amplitude of scalar perturbations $\Delta_S^2 \sim 10^{-10}$ (and $H^2 \approx \frac{1}{3M_{\rm pl}^2} V$) implies
\begin{equation}
V^{1/4} \sim \Bigl(\frac{r}{0.01} \Bigr)^{1/4} \, 10^{16}\, {\rm GeV}\, .
\end{equation}
Observable tensors therefore imply that inflation occurred at very high energies.
\item  Super-Planckian Field Variation\\
There is a one-to-one correspondence between the tensor-to-scalar ratio $r$ and the evolution of the inflaton in terms of the number of $e$-folds of inflation
\begin{equation}
\label{equ:rphi}
r = 8 \Bigl( \frac{1}{M_{\rm pl}} \frac{\dd \phi}{\dd N_e} \Bigr)^2\, .
\end{equation}
Integrating this from the end of inflation to 60 $e$-folds earlier, (\ref{equ:rphi}) implies the Lyth bound \cite{Lyth}
\begin{equation}
\frac{\Delta \phi}{M_{\rm pl}} > {\cal O}(1)  \Bigl(\frac{r}{0.01} \Bigr)^{1/2} \, .
\end{equation}
Observable gravitational waves therefore require $\Delta \phi > M_{\rm pl}$ while keeping the potential controllably flat.  Is this realizable in a consistent microscopic theory like string theory? (see {\it e.g.}~\cite{BM} for a discussion).
\end{enumerate}

\subsection{Final Remarks}
Cosmological observations are, for the first time, precise enough to test theoretical ideas about the early universe.
This has established inflation as the leading candidate for explaining both the large-scale homogeneity of the universe as well as the observed small-scale fluctuations.
In this lecture I gave a pedagogical introduction to inflationary cosmology with a special focus on the quantum generation of cosmological perturbations.
I have sacrificed many important technical details for a clear exposition of the physical principles behind the idea, but I hope that these notes may encourage the reader to teach herself the full story \cite{Mukhanov, Dodelson, LiddleLyth, MFB, proper}.





\vspace{1cm}
{\bf Acknowledgments:} I wish to thank Paul Steinhardt, Latham Boyle and Liam McAllister for numerous discussions on inflation which have contributed much to my understanding of the subject.
I am grateful to Malcolm Longair for convincing me that these notes might be useful to some people and for pushing me to make these notes publicly available.
Finally, I thank the theory groups at Caltech and Stanford for their kind hospitality while this work was completed.
\newpage
\appendix

\section{Classification of Single-Field Inflation Models}
\setcounter{equation}{0}
\renewcommand{\theequation}{A.\arabic{equation}}

In this appendix I present a useful classification of single-field inflation models through the curvature-slope ratio of the inflaton potential, $\alpha \equiv \eta/ \epsilon$. 
(A similar classification was introduced in Ref.~\cite{Kinney}).
Particular emphasis is put on predictions for the tensor-to-scalar ratio $r$ and the scalar spectral index $n_S$.\\

\noindent
{\sl First Order Slow Roll Analysis}. 

Recall the slow-roll parameters
\begin{eqnarray}
\epsilon &\equiv& \frac{1}{2}  \left( \frac{V'}{V}\right)^2  \\
\eta &\equiv&  \frac{V''}{V}
\end{eqnarray}
and their relation to the cosmological observables
\begin{eqnarray}
\Delta_S^2 &=& \frac{1}{24 \pi^2} \frac{V}{\epsilon} \\
n_S-1 &=& 2\eta -6 \epsilon  \\
r &=& 16 \epsilon  \, .
\end{eqnarray}
To assess the generic predictions of different classes of models for the tensor-to-scalar ratio $r$ it proves convenient to define the parameter
\beq
\alpha \equiv \frac{\eta}{\epsilon} = 2 \left( \frac{V''}{V'} \right) \left( \frac{V}{V'} \right)\, .
\eeq
You may easily verify that
\beq
\fbox{$\displaystyle
r = \frac{8}{\alpha-3} (n_S-1)$}\, .
\eeq
This suggests the following logic: Once future observation ({\it e.g.} by the Planck satellite) determine $n_S$ to an accuracy of $\Delta n_S = {\cal O}(0.01)$, the prediction for $r$ only depends on the $\alpha$ parameter of the model in question.\\

\noindent
{\sl Classification of Single-Field Inflation}. 

In the following we classify single-field inflation models on the basis of the $\alpha$--parameter.
\begin{itemize}
\item Class {\bf A} ($-\infty < \alpha < 0$): \\
These are models with negative curvature, $\eta < 0$:
The predictions for the spectral tilt and the tensor-to-scalar ratio are
\begin{equation}
n_S < 1\, , \quad \quad
0 \le r < \frac{8}{3} (1-n_S)\, .
\end{equation}
A generic potential for this class of models is
\beq
V(\phi) = \Lambda^4 \left[1-\left(\frac{\phi}{\mu}\right)^p \right] \approx \Lambda^4 \quad \quad p \ge 2, \quad \phi < \mu \, .
\eeq
\item Class {\bf B} ($0 < \alpha \le 2$):\\ 
These are models with
small positive curvature, $0 < \eta \le 2 \epsilon$:
Their predictions are
\begin{equation}
n_S < 1\, , \quad \quad
\frac{8}{3} (1-n_S) <  r \le 8 (1-n_S)\, ,
\end{equation}
and the generic potential is
\beq
V(\phi) = \Lambda^4 \left( \frac{\phi}{\mu} \right)^p\, \quad \quad p>1, \quad \phi \ll \mu \, .
\eeq

\item Class {\bf C} ($2 < \alpha \le 3$):\\
For these intermediate positive curvature models, $2\epsilon < \eta \le 3 \epsilon$,
the
predictions are
\begin{equation}
n_S < 1\, , \quad \quad
r > 8 (1-n_S)\, . 
\end{equation}

\item Class {\bf D} ($3 < \alpha < +\infty$):\\
These are large positive curvature models, $\eta > 3 \epsilon$, with
\begin{equation}
n_S \ge 1\, , \quad \quad
r \ge 0 \, .
\end{equation}
and
\beq
V(\phi) = \Lambda^4 \left[1+ \left( \frac{\phi}{\mu} \right)^p\right] \approx \Lambda^4  \quad \quad \phi < 1  \, .
\eeq
\end{itemize}

\begin{figure}[h]
	\centering
		\includegraphics[width=9cm]{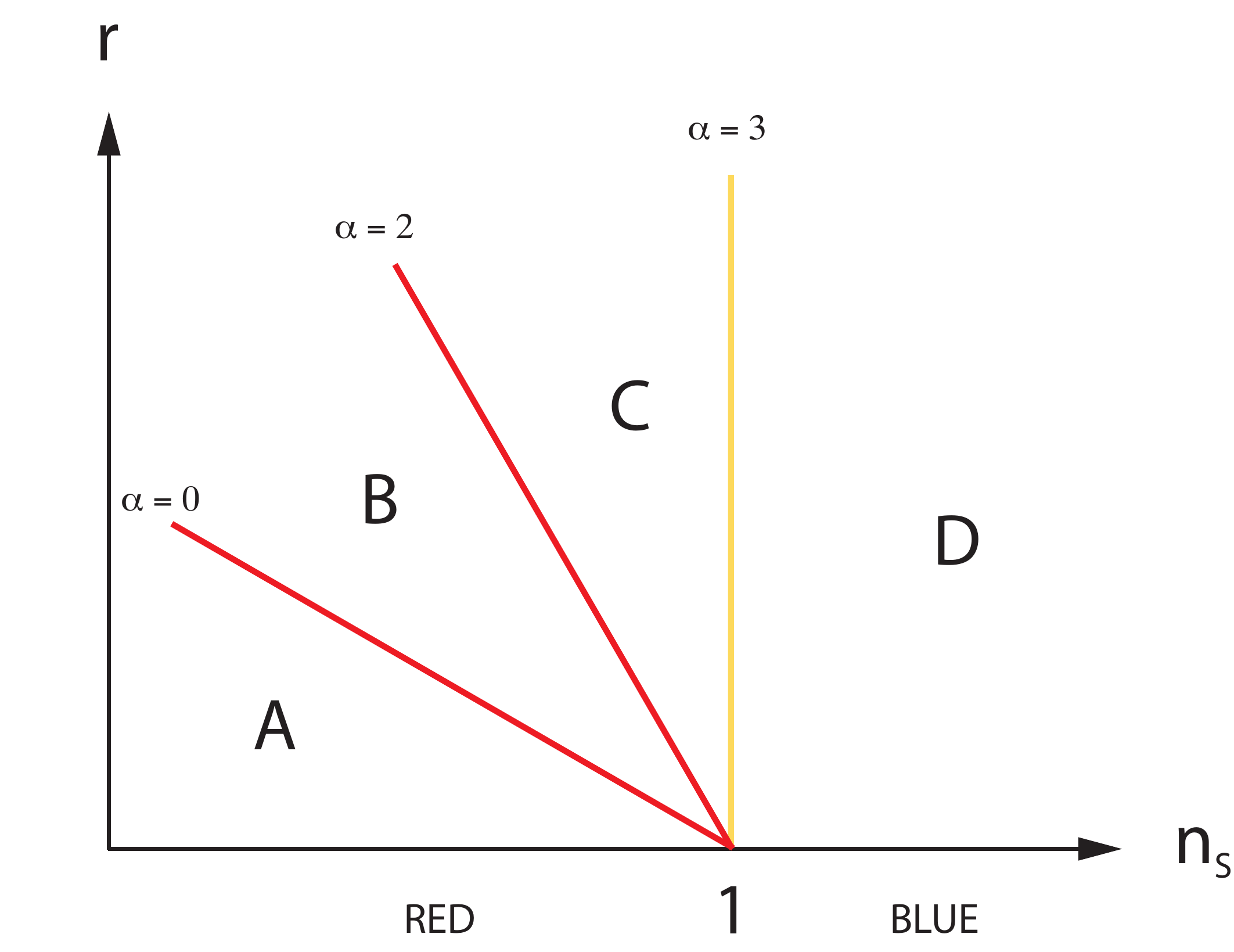}
	\caption{Future experiments will constrain inflationary models in the $n_S-r$ plane.}
	\label{fig:tilt}
\end{figure}

The $\alpha$--classification divides the $n_S-r$ plane into non-overlapping regions (Figure \ref{fig:tilt}).

\section{Slow-Roll Inflation in the Hamilton-Jacobi Approach}
\setcounter{equation}{0}
\renewcommand{\theequation}{B.\arabic{equation}}

\subsection{Hamilton-Jacobi Formalism}
The Hamilton-Jacobi approach treats the Hubble expansion rate $H(\phi)= {\cal H}/a$ as the fundamental quantity, considered as a function of time.
Consider
\beq
H_{,\phi} = \frac{H'}{\phi'} = \frac{-\frac{1}{a} ({\cal H}^2 -{\cal H}')}{\phi'} = - \frac{\phi'}{2a}\, ,
\eeq
where we used ${\cal H}^2 -{\cal H}' = a^2 (\rho + P)/2 =  (\phi')^2/2$ and primes are derivatives with respect to conformal time. This gives the master equation
\beq
\label{master}
\fbox{$\displaystyle
\frac{\dd \phi}{\dd t} = \frac{\phi'}{a} = -2 H_{,\phi}$}\, .
\eeq
This allows us to rewrite the Friedmann equation
\beq
H^2 = \frac{1}{3} \left[\frac{1}{2} \left(\frac{\dd \phi}{\dd t}\right)^2 +V(\phi)\right]
\eeq
in the following way
\beq
\label{HJ}
\fbox{$\displaystyle
[H_{,\phi}]^2 -\frac{3}{2} H^2 = \frac{1}{2} V(\phi)$}\, .
\eeq
Notice the following important consequence of the Hamilton-Jacobi equation (\ref{HJ}): For any specified function $H(\phi)$, it produces a potential $V(\phi)$ which admits the given $H(\phi)$ as an exact inflationary solution.
Integrating (\ref{master})
\beq
\int \dd t = -\frac{1}{2} \int \frac{\dd \phi}{ H'(\phi)}
\eeq
relates $\phi$ to proper time $t$. This enables us to obtain $H(t)$, which can be integrated to give $a(t)$. The Hamilton-Jacobi formalism can be used to generate infinitely many inflationary models with exactly known analytic solutions for the background expansion. Here we are more concerned with the fact that it allows an elegant and intuitive definition of the slow-roll parameters. 

\subsection{Hubble Slow-Roll Parameters}

During slow-roll inflation the background spacetime is approximately de Sitter. Any deviation of the background equation of state 
\[
w=\frac{P}{\rho} = \frac{(\phi')^2/2a^2 -V}{(\phi')^2/2a^2 + V}
\]
from the perfect de Sitter limit $w=-1$ may be defined by the parameter
\beq
\fbox{$\displaystyle
\epsilon_H \equiv \frac{3}{2}(1+w) $}\, .
\eeq 
We can express the Friedmann equations
\begin{eqnarray}
{\cal H}^2 &=& \frac{1}{3} a^2 \rho\\
{\cal H}' &=& -\frac{1}{6} a^2 (\rho+3p)
\end{eqnarray}
in terms of $\epsilon_H$
\begin{eqnarray}
{\cal H}^2 &=& \frac{1}{3} \frac{(\phi')^2}{\epsilon_H}\\
{\cal H}' &=& {\cal H}^2 (1-\epsilon_H) \, .
\end{eqnarray}
Hence,
\beq
\epsilon_H = 1-\frac{{\cal H}'}{{\cal H}^2} = \frac{\dd (H^{-1})}{\dd t}\, .
\eeq
Note that this can be interpreted as the rate ot change of the Hubble parameter during inflation $H$ with respect to the number of $e$-foldings $\dd N = H \dd t = -\frac{1}{2} \frac{H(\phi)}{H_{,\phi}} \dd\phi$
\beq
\fbox{$\displaystyle
\epsilon_H = -\frac{\dd [\ln H]}{\dd N} = 2 \left(\frac{H_{,\phi}}{H}\right)^2 $}\, .
\eeq
Analogously we define the second slow-roll parameter as the rate of change of $H_{,\phi}$
\beq
\fbox{$\displaystyle
\eta_H = - \frac{\dd [\ln |H_{,\phi}|]}{\dd N} = 2 \frac{H_{,\phi \phi}}{H}$}\, 
\eeq
Using the Hamilton-Jacobi master equation (\ref{equ:master}) this is also
\beq
\eta_H = \frac{\dd [\ln | \dd\phi/ \dd t|]}{\dd N}\, .
\eeq

\subsection{Slow-Roll Inflation}

By definition, slow-roll corresponds to a regime where all dynamical characteristics of the universe, measured in physical (proper) units, change little over a single $e$-folding of expansion. This ensures that the primordial perturbations are generated with approximately equal power on all scales, leading to a scale-invariant perturbation spectrum.

Since $\epsilon_H$ and $\eta_H$ characterize the rate of change of $H$ and $H_{,\phi}$ with $e$-foldings, slow-roll is naturally defined by
\begin{eqnarray}
\epsilon_H &\ll& 1\\
|\eta_H| &\ll& 1 \, .
\end{eqnarray}
The first slow-roll condition implies
\beq
\epsilon_H \ll 1 \quad \Rightarrow \quad {\cal H}^2 = \frac{1}{3} \frac{(\phi')^2}{\epsilon_H} \gg (\phi')^2\, ,
\eeq
so that the slow-roll limit of the first Friedmann equation is
\beq
{\cal H}^2 \approx \frac{1}{3} a^2 V \, .
\eeq
The second slow-roll condition implies
\beq
\eta_H = \frac{\dd [\ln |\dd\phi/\dd t|]}{\dd N} = \frac{1}{H |\dd\phi/\dd t|} \frac{\dd^2 \phi}{\dd t^2} \ll 1 \quad \quad \Rightarrow \quad \quad |\dd^2 \phi / \dd t^2| \ll H |\dd \phi /\dd t|
\eeq
so that the Klein-Gordon equation reduces to
\beq
\dot{\phi} \approx - \frac{a^2 V'}{3 {\cal H}}\, .
\eeq
In section \ref{sec:SR} we defined
a second set of common slow-roll parameters in terms of the local shape of the potential $V(\phi)$
\begin{eqnarray}
\epsilon &\equiv& \frac{1}{2} \left(\frac{V_{,\phi}}{V}\right)^2\\
\eta &\equiv& \frac{V_{,\phi \phi}}{V}\, .
\end{eqnarray}
$\epsilon_H(\phi_{\rm end}) \equiv 1$ is an exact definition of the end of inflation, while $\epsilon(\phi_{\rm end}) =1$ is only an approximation. 
In the slow-roll regime the following relations hold
\begin{eqnarray}
\epsilon_H &\approx& \epsilon \\
\eta_H &\approx& \eta - \epsilon\, .
\end{eqnarray}

\end{document}